\documentclass[
	12pt
]{iopart}
\usepackage{graphicx}
\usepackage[tight]{subfigure}
\usepackage{amsmath}
\usepackage[sort&compress,square,numbers]{natbib}
\usepackage{dcolumn}% Align table columns on decimal point
\usepackage{bm}% bold math
\begin{document}

\renewcommand{\thefigure}{\arabic{figure}}
\newcommand{\ome}{\omega}                 % omega
\newcommand{\omen}{\omega_{ne}^{\ast}}      % Density diamagnetic frequency
\newcommand{\omepe}{\omega_{pe}^{\ast}}      % Pressure diamagnetic
\newcommand{\omepi}{\omega_{pi}^{\ast}}      % Pressure diamagnetic
\newcommand{\omeu}{\omega_{u}^{\ast}}      % Velocity diamagnetic frequency
\newcommand{\omed}{\omega_d}              % Vertical drift frequency
\newcommand{\kperp}{k_{\perp}}              % kperp
\newcommand{\kpar}{k_{\parallel}}           % kparallel
\newcommand{\kthe}{k_{\theta}}              % ktheta
\newcommand{\QLK}{QuaLiKiz}
\newcommand{\GKW}{\textsc{gkw}{}}
\newcommand{\Gene}{\textsc{gene}}
\newcommand{\gst}{\textsc{gs}2}
\newcommand{\gyro}{\textsc{gyro}}
\newcommand{\vpar}{v_\parallel}           % vparallel
\newcommand{\upar}{u_\parallel}           % uparallel
\newcommand{\dt}{\frac{\partial}{\partial t}}
\newcommand{\dtheta}{\mathrm{d}\theta}
\newcommand{\dx}{\frac{\mathrm{d}}{\mathrm{d}x}}
\newcommand{\dr}{\frac{\mathrm{d}}{\mathrm{d}r}}
\newcommand{\ddx}{\frac{\mathrm{d}^2}{\mathrm{d}x^2}}
\newcommand{\ds}{\displaystyle}
\newcommand{\deff}{d_{\textsl{eff}}}
\newcommand{\ceff}{c_{\textsl{eff}}}
\newcommand{\reff}{\rho_{\textsl{eff}}}
\newcommand{\w}{\mathrm{w}}
\newcommand{\x}{\mathrm{x}}
\newcommand{\n}{\mathbf{n}}
\renewcommand{\d}{\mathrm{d}}
\newcommand{\J}{\mathbf{J}}
\newcommand{\I}{\mathcal{I}}
\newcommand{\OmeJ}{{\n\cdot\boldsymbol{\Omega}_\mathbf{J}}}  % omega_J
\newcommand{\Omestar}{\n\cdot{\boldsymbol{\Omega^\ast}}}     % omega*
\newcommand{\Omestars}{{\n\cdot\boldsymbol{\Omega}}_s^\mathbf{\ast}}     % omega*_s
\newcommand{\gradu}{\nabla{u_\parallel}}
\newcommand{\gradun}{\frac{R\gradu}{v_{Ti}}}
\newcommand{\zeff}{Z_{\textsl{eff}}}
\newcommand{\gameff}{\gamma_{\textsl{eff}}}
\newcommand{\uparn}{\frac{\upar}{v_{Ti}}}
\newcommand{\uparns}{\frac{\upar}{v_{Ts}}}
\newcommand{\exb}{\mathbf{E}\times\mathbf{B}}
\newcommand{\omE}{\omega_{\exb}} %\omega_ExB 

\title{Angular momentum transport modeling: achievements of a gyrokinetic quasi-linear approach}

\author{P~Cottier$^{1}$, C~Bourdelle$^1$, Y~Camenen$^2$, \"{O}~D~G{\"u}rcan$^3$, F~J~Casson$^4$, X~Garbet$^1$, P~Hennequin$^2$ and T~Tala$^5$}
%\email{pierre.cottier@cea.fr}
%\author{C.~Bourdelle}
\address{IRFM, CEA, F-13108 Saint Paul-Lez-Durance, France}
\ead{pierre.cottier@cea.fr}
%\author{Y.~Camenen}
\address{PIIM UMR 7345, CNRS, Aix-Marseille Univ., Marseille, France}
%\author{\"{O}.D.~G{\"u}rcan}
\address{LPP UMR 7648, CNRS-Ecole Polytechnique, Palaiseau, France}
%\author{F.~Casson}
\address{Max-Planck-Institut f{\"u}r Plasmaphysik, IPP-Euratom Association, Garching-bei-M{\"u}nchen, Germany}
%\author{X.~Garbet}
%\affiliation{IRFM, Association Euratom-CEA, Saint Paul Lez Durance, France}
%\author{P.~Hennequin}
%\affiliation{LPP, UMR 7648 CNRS-Ecole Polytechnique, Palaiseau, France}
%\author{T.~Tala}
\address{VTT, Association Euratom-Tekes, PO Box 1000, FIN-02044 VTT, Finland}
\begin{abstract}
\QLK, a model based on a local gyrokinetic eigenvalue solver\citep{bour02} is expanded to include momentum flux modeling in addition to heat and particle fluxes\citep{bour07,casati09}. Essential for accurate momentum flux predictions, the parallel asymmetrization of the eigenfunctions is successfully recovered by an analytical fluid model. This is tested against self-consistent gyro-kinetic calculations and allows for a correct prediction of the $\exb$ shear impact on the saturated potential amplitude by means of a mixing length rule. Hence, the effect of the $\exb$ shear is recovered on all the transport channels including the induced residual stress. Including these additions, \QLK{} remains $\sim 10000$ faster than non-linear gyro-kinetic codes allowing for comparisons with experiments without resorting to High Performance Computing. The example is given of momentum pinch calculations in NBI modulation experiments\citep{tala09} for which the inward convection of the momentum is correctly predicted. 
\end{abstract}

\pacs{52.25.Fi, 52.30.Gz, 52.35.-g, 52.35.Ra, 52.55.Fa, 52.65.-y, 52.65.Tt, 52.65.Vv}

\maketitle

\section{Introduction}
\label{sec:intro}
Sheared flows in tokamaks have long been studied since there are both theoretical and experimental evidences that they can significantly enhance the plasma energy confinement \citep{biglari90,ida90,stambaugh90,burrel97,devries08}. The toroidal torque can result from the interaction of the turbulent plasma with the walls and the coils \citep{Rice04,abiteboul13,Fenzi11} or from the heating system such as NBI \citep{groebner90,scott90,honda09} or even RF heating\citep{eriksson01}. The back-reaction of sheared flows on turbulence has received considerable attention, either its stabilizing effect with sheared \emph{poloidal} rotation (related to sheared radial electric field via the $\exb$ drift) \citep{biglari90,burrel97} or its destabilizing effect with parallel velocity gradient $\gradu$\citep{dangelo65,artun92,peet05}. The interplay between mean flows and turbulence can be described quantitatively by quasilinear fluid models\citep{staebler05,kinsey08,staeb13}, non-linear gyro-fluid models\citep{garb96,waltz98}, quasilinear gyrokinetic models\citep{angioni11,camenen09,tala09}, and non-linear gyrokinetic simulations\citep{barnes11,highco10,kinsey05,peet11,roach09,waltz07,strugarek13}. 

This paper presents a reduced model compatible with integrated modeling able to predict both momentum transport and sheared flows effects on turbulence for tokamak plasmas. This model is extending the \QLK{} transport code abilities which was developed to compute heat and particle fluxes \citep{bour07}. The philosophy of \QLK{} is to minimize the number of \emph{ad hoc} parameters. Only the saturated potential amplitude is prescribed once and for all to match the ion heat flux of non-linear gyrokinetic simulations for the GA-std case.  Predicting quantitatively the turbulent fluxes without resorting to parameter fitting requires the use of a gyrokinetic linear solver. However two orders of magnitude in CPU time have to be gained to be compatible with the integrated modeling framework. Therefore, \QLK{} uses both the ballooning representation at lowest order, reducing the dimension of the problem to 3 from ($\mu$,$\vpar$,$r$,$\theta$) to ($\mu$,$\vpar$,$r$) by a Fourier decomposition in the radial direction, and trial eigenfunctions from the analytic fluid limit \citep{bour02,roma07}. \QLK{} is coupled to CRONOS, an integrated modeling platform that evolves consistently $q$, $T_e$, $T_i$ and $n_e$ profiles\cite{artaud09}. It has been used for the prediction of the heat transport in JET \cite{Baiocchi13}. 

In the new version of \QLK, the impact of the plasma rotation on the eigenfunction is reproduced with satisfactory accuracy compared to self-consistent gyrokinetic codes through a complex shift of Gaussian eigenfunctions. The effect of this shift is included in the non-linear saturation rule through the use of an effective $k_\bot$ as detailed in Sec.~\ref{sec:fluxes}. However different from the model proposed in \citep{staeb13}, the method detailed in this work enables the recovery of the heat and particle flux stabilization with $\exb$ shear. The induced residual stress can also be estimated with the benefit to be fitting-parameter free. However, the local approach taken in \QLK{} does not allow for a consistent treatment of higher $\rho^*$ effects characterizing the residual stress \citep{waltz11,camenen11,gurcan10c}. The heat and particle flux reduction with $\exb$ shear match non-linear gyrokinetic simulations. The momentum flux sensitivity to $\upar$ and $\gradu$ is also in agreement with non-linear gyrokinetic results and shows the importance to have the correct shape of the eigenfunctions in the parallel direction. Finally, the comparison with NBI modulation experiments showing the existence of an inward convective momentum flux in JET\citep{tala09} is successful. It underlines that separating the different contributions to the momentum flux is challenging. The fluxes sensitivity to the gradient estimations is highlighted, advocating for flux forcing of the code, which is to be done by coupling this new version to CRONOS integrated platform. 

First, the linear eigenfunction/eigenvalue equation at the heart of the linear solver of \QLK{} is re-derived in Sec.~\ref{sec:gyrok} to include new terms coming from the plasma bulk rotation. Then, in Sec.~\ref{sec:fluid}, the fluid model calculating the eigenfunctions is revisited to include the sheared flow effects and compared to self-consistent gyrokinetic eigenfunctions from the gyrokinetic code \GKW\citep{peetCPC09}. In Sec.\ref{sec:growth rates}, the sensitivity to $\upar$, $\gradu$ and $\exb$ shear of \QLK{} linear growth rates are successfully benchmarked against \GKW. In Sec.~\ref{sec:fluxes}, the quasi-linear momentum flux is derived, the shape of the saturated potential is discussed and the estimations of heat, particle and momentum fluxes are compared to non-linear gyrokinetic simulations from \GKW{} and \gyro{}. The methods to separate the different contributions to the momentum flux are discussed as well. Finally, in Sec.~\ref{sec:exp}, a JET shot with NBI modulation \citep{tala07} is modeled. The diffusive and convective terms are compared to the experimental values. 
%Using the new abilities of \QLK{}, it is shown that the residual stress can be significant in these experiments.

\section{Linearized gyrokinetic dispersion relation}
\label{sec:gyrok}
First, the linearized gyrokinetic equation is derived including the effect of a finite rotation of the plasma. The formalism employed in previous derivations without bulk rotation \citep{bour02} is conserved and its validity range is discussed. Finally, the expression used for the linear solver in \QLK{} and based on the linearized Vlasov equation coupled with the electroneutrality condition is given.

To study the impact of the plasma rotation, the model has to allow for a finite equilibrium rotation of the system $\upar$. In the gyrokinetic framework, this translates into having a finite value for $\upar=\int{f_0v_\parallel\d^3{v}}$, the integral of the product of $f_0$, the equilibrium distribution function multiplied by the velocity coordinate $\vpar$. $f_0$ being chosen Maxwellian, it reads, for each species $s$ of density $n_s$, mass $m_s$, temperature $T_s$ and thermal velocity $v_{Ts}=\sqrt{2T_s/m_s}$:
\begin{equation}
f_0^s=\frac{n_s}{(2\pi T_s/m_s)^{3/2}}\exp\left(-\frac{E}{T_s}+\frac{\upar(2\vpar-\upar)}{v_{Ts}^2}\right)
\end{equation}
$\vpar$ being the parallel velocity coordinate, $E$ the energy defined by $E/T_s=\vpar^2/v_{Ts}^2+\mu B/T_s$ and $\mu$ the adiabatic invariant. The reference frame being the laboratory frame here $\frac{\upar}{v_{Ts}}$ is the Mach number for the species $s$. In core plasma of conventional tokamaks, the Mach number is usually limited to values smaller than 0.4. In spherical tokamaks, however, core Mach numbers can reach values close to unity\cite{roach09}. The low Mach number limit is taken in the following and $f_0$ is developed up to second order in $\frac{\upar}{v_{Ts}}$.
Now, taking the linearized Vlasov equation in the angle-action variables $(\boldsymbol\alpha,\J)$ and applying quasi-neutrality in its variational form\citep{garb90}, one finds\citep{bour02}:
\begin{equation}
\sum_{s}\left\langle\frac{\n\cdot\partial_\J{f_0^s}}{\omega-\n\cdot\partial_\J h_0+\imath o^+}|\tilde h_{\n,\omega}|^2\right\rangle_{\J,{\boldsymbol\alpha}}=0
\label{eq:vlasov}
\end{equation}
where $\J$ are the actions i.e. the three invariants: $\mu$, $E$ and $p_\phi$ the angular momentum. $\boldsymbol{\alpha}$ are the associated angles defined by $\boldsymbol{\dot\alpha}=\partial_\J h_0$. $\n$ are the wave numbers associated with the angle variables $\boldsymbol{\alpha}$ through the Fourier decomposition of the fluctuating distribution function and fluctuating Hamiltonian\citep{garb90}. $h_0$ is the unperturbed Hamiltonian defined by $h_0= mv^2/2+e\phi$. \QLK{} is an electrostatic code, the unperturbed Hamiltonian being reduced to its electrostatic part. The brackets $\langle\cdots\rangle_{\J,{\boldsymbol\alpha}}$ mean integration over $\J$ and $\boldsymbol{\alpha}$. See Appendices A.1 and A.2 of \citep{bour02} for a detailed derivation of the linearized Vlasov equation, its decomposition over the angle-action variables and how the electroneutrality condition is used to find (\ref{eq:vlasov}). In the electrostatic limit, the perturbed Hamiltonian $\tilde{h}_{n\omega}$ is reduced to $e_s\tilde\phi_{n\omega}$. It is clear from this equation that terms proportional to the parallel velocity and its gradient, coming from $\n\cdot\partial_\J{f_0}$, will impact the linear response. To illustrate this, the diamagnetic frequency $\Omestar=\n\cdot\partial_\J{f_0}-\n\cdot\partial_\J{E}/T_s$ is expressed as a function of the gradients $\nabla{n_s}$, $\nabla{T_s}$ and $\nabla\upar$ in (\ref{eq:dia_freq}).
\begin{equation}
\label{eq:dia_freq}
\Omestars=\frac{\kthe T_s}{e_sB} \Big[\frac{1}{n_s}\frac{\d{n_s}}{\d{r}}+\left(\frac{E}{T_s}-\frac{3}{2}-\frac{\upar}{v_{Ts}}\frac{2\vpar-\upar}{v_{Ts}}\right)\frac{1}{T_s}\frac{\d{T_s}}{\d{r}}+ 2\left(\frac{\vpar-\upar}{v_{Ts}}\right)\frac{1}{v_{Ts}}\frac{\d\upar}{\d{r}}\Big]
\end{equation}
where $\kthe=\frac{-nq}{r}$ is the poloidal wave vector in the ballooning representation presented later on, $n$ being the toroidal wave number, $q$ the safety factor and $r$ the radial coordinate. $e_s$ is the charge of the species $s$. 
%Let us introduce the normalized gradients: $A_{ns}=-R\nabla{n_s}/{n_s}$, $\frac{R}{L_{Ts}}=-R\nabla{T_s}/{T_s}$, $A_{u}=-R\nabla{\upar}/{v_{Ts}}$ and the quantity $n\bar\omega_{ds}=-\frac{\kthe T_s}{e_sBR}$ which is physically defined below. With these definitions, (\ref{eq:dia_freq}) becomes:
%\begin{equation}
%\Omestars=n\bar\omega_{ds}\left(A_{ns}+\left(\frac{E}{T_s}-\frac{3}{2}-\frac{\upar}{v_{Ts}}\frac{2\vpar-\upar}{v_{Ts}}\right)\frac{R}{L_{Ts}}+ 2\left(\frac{\vpar-\upar}{v_{Ts}}\right)A_u\right)
%\end{equation}
%illustrating that the diamagnetic frequency is proportional to the bulk velocity and its gradient.

From (\ref{eq:vlasov}), $\n\partial_\J h_0$ corresponds to the three frequencies associated with the three angle variables ($\n\partial_\J h_0=\partial_t\boldsymbol{\alpha}$) characterizing the particles movement within the magnetic field of a tokamak, namely the cyclotron frequency, the parallel motion frequency (bounce frequency for trapped particles) and the -- curvature, $\nabla{B}$ and $\exb$ -- drift frequency. Since the cyclotron frequency $\omega_c$ is much larger than the other characteristic frequencies, a scale separation is possible. The dependence over the gyro-angle can be removed either by averaging over the gyromotion according to historical gyrokinetic theory \citep{rutherford68,taylor68,antonsen80,catto81} or via Lie transforms according to modern gyrokinetics\citep{littlejohn81,hahm88b,brizard89b,brizard07}. In the end, both methods result in multiplying the perturbed potential $\tilde{h}$ by the zero order Bessel function $J_0^2(k_\bot\rho_s)$ ; $\rho_s$ being the Larmor radius for the species $s$. Then $\n\cdot\partial_\J h_0=\OmeJ$ corresponds to the gyrocenter drifts. In the simplified $\hat{s}-\alpha$ equilibrium, which \QLK{} is using, $\OmeJ$ can be written as expressed in (\ref{eq:drifts}). Using such an equilibrium leads to the underestimation of ITG linear growth rates with respect to more consistent circular magnetic equilibria as shown in \citep{lapillonne09} (see Figure~6 from \citep{cass10} too).
\begin{equation}
\label{eq:drifts}
\OmeJ=n\omega_{ds}+n\omE+\kpar\vpar=-(2-\lambda b)\frac{\kthe T_s}{e_sBR}(\cos\theta+(\hat{s}\theta-\alpha\sin\theta)\sin\theta)\frac{E}{T_s}+\frac{\kthe E_r}{B}+\kpar\vpar
\end{equation}
$\lambda=\frac{\mu B}{E}$ is the pitch-angle and $b(r,\theta)=\frac{B(r,\theta)}{B(r,0)}$ is the magnetic field normalized to its value at the outboard midplane. $\hat{s}$ is the magnetic shear and $\alpha=-q^2\beta\nabla{P}/P$ is the MHD parameter.

The first term, $n\omega_{ds}$, in (\ref{eq:drifts}) corresponds to the curvature and $\nabla{B}$ drifts whose expression is valid only in the low $\beta$ limit \citep{bour03} and at lowest order in $\epsilon$\citep{lapillonne09}. The second term, $n\omega_{\exb}=\frac{\kthe E_r}{B}$, is the $\exb$ drift and the last term is associated with the fast parallel motion of particles. $\kpar\vpar$ expression is given by (\ref{eq:kparvpar}).
\begin{equation}
\label{eq:kparvpar}
\kpar\vpar=\pm\frac{v_{Ts}x}{qRd}\sqrt{\xi(1-\lambda b)}
\end{equation}
where $d=\frac{1}{\kthe\hat{s}}$ is the distance between resonant surfaces such that $q=m/n$ and $x$, the distance to the closest resonant surface. 
The curvature and $\nabla{B}$ drift is expressed as $n\omega_{ds}=(2-\lambda b)n\bar\omega_{ds}f_\theta\xi$ with $f_\theta=\cos\theta+(\hat{s}\theta-\alpha\sin\theta)\sin\theta$ and $\xi=E/T_s$.
Overall, (\ref{eq:vlasov}) reads:
\begin{equation}
\label{eq:electroneutralite}
\sum_{s}{\frac{e_s^2n_s}{T_s}}\left\langle\left(1+\frac{2\upar\vpar}{v_{Ts}^2}+\frac{\upar^2}{v_{Ts}^2}\left(\frac{2\vpar^2}{v_{Ts}^2}-1\right)\right)e^{-\xi}\left(1-\frac{\omega-n\omega_{E\times B}-\Omestars}{\omega-\OmeJ+\imath o^+}\right)J_0^2(\kperp\rho_s)\left|\tilde\phi_{n\omega}\right|^2\right\rangle=0
\end{equation}
$n\omE=k_\theta E_r/B$ appearing in (\ref{eq:electroneutralite}) results from the simplification of $\n\cdot\partial_\J{E}/T_s$ from $\n\cdot\partial_\J{f_0}$ with $\n\cdot\partial_\J{h_0}$. $n\omE$ is species independent so $\omega-n\omega_{E\times B}$ can be replaced by one variable $\varpi$.
 
One important approximation made in \QLK{} is the use of the ballooning representation \citep{connor78,pegoraro81,dewar83} truncated at lowest order i.e. only the lowest harmonic in the infinite sum is retained. In this case, the ballooning representation comes down to an infinite sum of identical modes at $(r_0,n)$ position, $r_0$ being a resonant $q=m/n$ radius (\citep[see][App. A.1]{bour02} or \citep{candy04}). This enables a local treatment in $r$ at the expense of a limitation on the $\theta$ expansion of the mode to $\theta\in[-\pi; \pi]$ as illustrated in \ref{app:TEM}. The integration over $\J$ and ${\boldsymbol\alpha}$ then comes down to integration over the pitch-angle, the energy $\xi=\frac{E}{T_s}$ and $k_r$. Indeed, $\theta$ integration is done through $\theta=k_rd$ \citep{pegoraro81} and axisymmetry allows for Fourier decomposition in the toroidal direction. At this stage, it is important to acknowledge that such an approximation is valid only if the eigenmodes are sufficiently coupled together by the magnetic shear. A condition for that is the mode width $\w$ -- expressed in Sec.~\ref{subsec:fluid model} -- to be much larger than $d$. This is equivalent to say the eigenfunction is peaked and does not expand outside $[-\pi; \pi]$. It was validated down to $\hat{s}=0.1$ and $\kthe\rho_s=0.15$ \citep[see][App. C]{citrin12}. In addition, the gradient lengths $L_x$ (among density, temperature, velocity, safety factor) must satisfy:
\begin{equation}
\label{eq:local_scale}
L_x\gg d
\end{equation}
to ensure that the envelope effects are small. For highly sheared plasma flows, the validity of the approach has to be considered. If the velocity gradient scale length is defined as $L_u=\frac{v_{Ts}}{\gradu}$, it can reach values as small as $R/5$ in core tokamak plasmas. The condition (\ref{eq:local_scale}) then becomes $\epsilon\ll nq\hat{s}/5$ where $\epsilon$ is the inverse aspect ratio. So, for highly rotating plasmas, the approach is valid down to $\hat{s}\geq0.2$ and $n\geq10$ which is similar to the limitations seen in \citep{citrin12}. The issue of the ballooning representation compatibility with a poloidal sheared velocity has been extensively studied\citep{cooper88,miller94,waltz98}. Nevertheless, since (\ref{eq:local_scale}) is satisfied for experimental values of $L_{\gamma_E}=v_{Ts}/\gamma_E=v_{Ts}B/\frac{d{E_r}}{\d{r}}>R$, it is considered that modes remain ballooned around $\theta=0$ and the ballooning representation is used truncated at lowest order. 

In \QLK, the response of trapped and passing particles are separated to take advantage of their different dynamics. An average over the bounce motion is performed for trapped particles, reducing further the numerical cost of the model because it enables the removal of the $\theta$ dependence of the drift frequencies. In the same spirit as the gyromotion average, bounce motion average results in the multiplication of the trapped particles response by Bessel functions $J_m(k_r\delta_s)$, $k_\bot$ coming down to $k_r$ in the thin banana approximation. $\delta_s$ is the banana width of the species $s$. Because of the assumption of local Maxwellian equilibrium, the Bessel functions integration is done separately giving $\mathcal{B}_m(a)=\exp(-a^2)I_m(a^2)$ (\citep[see][App.A.4]{bour02} for the $m=0$ case). (\ref{eq:electroneutralite}) can be written under the condensed form (\ref{eq:electroneutralite2}), $\mathcal{I}_{s,m,tr}$ and $\mathcal{I}_{s,pass}$ expression being detailed in ~\ref{subsec:trapped} and \ref{subsec:passing} respectively.
\begin{equation}
\label{eq:electroneutralite2}
\sum_{s}{\frac{e_s^2n_s}{T_s}}\left[1-\int\frac{\d{k_r}}{2\pi}\left(\left\langle\mathcal{I}_{s,pass}\right\rangle_p\mathcal{B}_0(k_\perp\rho_s)-\sum_{m}\left\langle\mathcal{I}_{s,m,tr}\right\rangle_t\mathcal{B}_0(k_\perp\rho_s)\mathcal{B}_m(k_r\delta_s)\right)\right]=0
\end{equation}
The integration over the passing domain is
\begin{equation*}
\langle\cdots\rangle_p=\int_0^\infty\frac{2\sqrt{\xi}}{\sqrt{\pi}}\exp(-\xi)\d\xi\int_0^{\lambda_c}\frac{\d\lambda}{4\bar\omega_b}
\end{equation*}
$\lambda_c=\frac{1-\epsilon}{1+\epsilon}$ is the minimum value of the pitch angle for which particles can be trapped and $\bar\omega_b$ is the normalization of $\lambda$ over the parallel (or bounce) motion $\bar\omega_b^{-1}=\oint{\frac{\d\theta}{2\pi}\frac{1}{\sqrt{1-\lambda b}}}$ with $\oint=\int_\pi^\pi$ for passing particles and $\oint\approx2\int_{-\theta_b}^{\theta_b}$ for trapped particles, $\theta_b$ being the bouncing point of the trapped particles. The integration over the trapped domain then reads:
\begin{equation*}
\langle\cdots\rangle_t=\int_0^\infty\frac{2\sqrt{\xi}}{\sqrt{\pi}}\exp(-\xi)\d\xi\int_{\lambda_c}^1\frac{\d\lambda}{4\bar\omega_b}=f_t\int_0^\infty\frac{2\sqrt{\xi}}{\sqrt{\pi}}\exp(-\xi)\d\xi\int_0^1K(\kappa)\kappa\d\kappa
\end{equation*}
where $f_t$ is the fraction of trapped particles, $\kappa$ is related to the pitch-angle via $\lambda=1-2\epsilon\kappa^2$ and $K$ is the complete elliptic integral of the first kind. For the expression of $\mathcal{I}_{s,pass}$ and $\mathcal{I}_{s,tr}$ please refer to ~\ref{subsec:passing} and \ref{subsec:trapped} where the algebra is detailed. 

In short, with the definitions given in ~\ref{subsec:passing} and \ref{subsec:trapped}, the expression (\ref{eq:electroneutralite2}) can be written as
\begin{equation}
\label{eq:lin_gyroK}
\sum_s{\frac{n_se_s^2}{T_s}\left(1-\mathcal{L}_{s,pas}(\omega)-\mathcal{L}_{s,tr}(\omega)\right)}=0
\end{equation}

In this section, the linearized gyrokinetic dispersion relation (\ref{eq:lin_gyroK}) at the heart of the linear solver of \QLK{} was derived including the effect of the non-zero values for $\upar$, $\gradu$ and $\exb$ shift in the low Mach number approximation and other standard approximations for \QLK, namely low $\beta$ (electrostatic), large aspect ratio and lowest order ballooning representation. The detail of the various functionals is available in ~\ref{subsec:passing} and \ref{subsec:trapped}. To solve this eigenfunction/eigenvalue equation, the eigenfunction $\tilde\phi$ is calculated in the analytic fluid limit, which is revisited in the following section to include the effect of sheared flows.

\section{Analytic eigenfunction calculation}
\label{sec:fluid}
In \QLK{}, the eigenmodes are not self-consistently calculated from (\ref{eq:lin_gyroK}). To gain CPU time -- 2 orders of magnitude together with the dimension reduction associated with the ballooning approximation detailed in previous section -- they are calculated in the fluid limit in which (\ref{eq:lin_gyroK}) can be solved analytically. This method proved to give satisfactory results compared to self-consistent gyrokinetic calculations in the case with no rotation (see in particular \citep[][Appendix C]{citrin12} and \citep[][Appendix A]{roma07}). A model for analytic eigenfunctions in presence of sheared flows is derived in this section. It is shown that shifted Gaussians are satisfactory approximates of the gyrokinetic eigenfunctions in that case. A comparison against \GKW \citep{peetCPC09} is performed as a validation for the cases with rotation. 

Fluid modeling of the linear eigenmode equation to find an analytic solution for the eigenfunction is not a new idea\citep{artun92,diam88,dong93,garb02,gurcan07,hahm07}. Here, the derivation is performed in the toroidal geometry and include the effects from $\upar$, $\gradu$ and the $\exb$ shear. (\ref{eq:lin_gyroK}) being the starting point of this derivation, all previous approximations still apply in particular the low Mach number approximation and the ballooning representation at lowest order. 

\subsection{Description of the fluid model}
\label{subsec:fluid model}
The fluid limit approximation consists in considering events sufficiently fast decorrelated by collisions such that $\varpi=\omega-n\omega_{E\times B}\gg\bar\omega_{di}$ and $\varpi\gg\kpar v_{\parallel i}$. This approximation enables the development of the dispersion relation given in (\ref{eq:lin_gyroK})in power of the small quantities $\frac{\omega_{ds}}{\varpi}$, $\frac{\kpar \vpar}{\varpi}$ and obtain a polynomial expression in $\varpi$ as detailed in (\ref{eq:electroneutralite4}).

For short wavelengths: $k_\bot\rho_i<1$, the Bessel functions can be linearized such that $\mathcal{B}_0(k_\bot\rho_i)\approx 1-\frac{k_\bot^2\rho_i^2}{2}$. At this spatial scale, events are sufficiently slow such that $\omega\ll \kpar v_{\parallel e}$. Passing electrons are then considered adiabatic. In contrast, TEM space and time scales being the same as ions modes, trapped electrons are treated by the model. Since $k_r\delta_e<k_r\rho_i<1$, the Bessel functions on trapped electrons are considered close to unity $\mathcal{B}_0(k_r\delta_e)\approx 1$. For trapped ions, the finite banana width effects are expended in power of $k_r$ too: $ \mathcal{B}_0(k_r\delta_i)\approx 1-\frac{k_r^2\delta_i^2}{2}$. The resulting polynomial expression for the eigenmode is given in (\ref{eq:electroneutralite4}). 

As explained in detail in \ref{app:fluid}, the electroneutrality condition $\sum_s{e_sn_s}=0$ is used to reformulate (\ref{eq:electroneutralite4}). It enables a species independent normalization frequency $n\bar\omega_d=n\bar\omega_{de}=-T_e/T_in\bar\omega_{di}$.
An inverse Fourier transform $k_r\rightarrow -\imath\partial_x$ is performed leading to a second order differential equation. $\varpi$ is replaced with $\omega-n\omE$ because $n\omE$ has an $x$ dependence in case of $\exb$ shear. The radial electric field is considered smooth enough such that it can be linearized into $E_r\rightarrow E_{r0}+E_r'x+ O(x^2)$ implying the linearization of $n\omE$ in $x$: $n\omE=n\omega_{E0}+\kthe\gamma_E x+O(x^2)$. Therefore, only the linear terms in $\gamma_Ex$ are taken into account in the eigenmode equation. The details of the derivation of the eigenmode equation are detailed in \ref{app:fluid}. Its final expression is given by (\ref{eq:lim_fluide2}).
\begin{align}\label{eq:lim_fluide2}\begin{split}
\Bigg[&\left(\omega\left(\frac{\deff^2}{2}\ddx-\frac{\kthe^2\reff^2}{2}\right) +\frac{\kpar^{\prime 2}\ceff^2}{2\omega}x^2\right)\left(\omega-n\omepi\right) -2n\bar\omega_{d}(\omega-\kthe\gamma_E)
-\omega^2+2\kthe\gamma_E+\\&\left(\omega-\kthe\gamma_E\right)n\omen-\frac{f_t}{f_p}n\omepe n\bar\omega_{d}+ \kpar^\prime\ceff\left(n\omeu+\frac{\upar}{\ceff}\left(\frac{\zeff}{\tau}\omega+n\omen-8n\bar\omega_{d}\right)\right)x\Bigg]\tilde\phi=0
\end{split}\end{align}
The solution of this linear second order differential equation is a shifted Gaussian:
\begin{equation}
\tilde\phi=\frac{\phi_0}{\left(\pi\Re\left(\w^2\right)\right)^{1/4}}\exp-\frac{(x-\x_0)^2}{2\w^2}
\end{equation}
This solution is characterized by two quantities:
\begin{itemize}
\item The \emph{mode width} $\w$ determined by: $\ds{\w^2=\frac{-\imath\omega \deff}{|\kpar^\prime|\ceff}}$, $\omega$ being the self-consistent solution of (\ref{eq:lim_fluide2}). The mode width therefore depends on $\gamma_E$, $n\omeu$ and $\upar$ through $\omega$.  Note that $\w^2$ was previously calculated with an interchange ansatz for $\omega$ in \QLK{} considering $\w$ real, it is defined here to cancel the quadratic terms in $x$ in (\ref{eq:lim_fluide2});
\item The \emph{mode shift} $\x_0$ characterizing the parallel asymmetrization of the mode expressed by:
\end{itemize}
\begin{equation}
\label{eq:shift}
\x_0=\frac{2n\bar\omega_d}{\omega-\omen}\frac{\frac{q}{s}\gamma_E^N(2\omega+2n\bar\omega_d-n\omen)+n\omeu+\frac{\upar}{\ceff}\left(\frac{\zeff}{\tau}\omega+n\omen-8n\bar\omega_{d}\right)}{\kpar^\prime\ceff}
\end{equation}
where $\gamma_E^N=\frac{\gamma_E}{\ceff/R}$ corresponds to usual normalizations of the $E\times B$ shear. The approach taken here to include consistently the effect of the $E\times B$ shear in the linear eigenfunctions is quite different than what is used in GLF 23/TGLF \citep{waltz97,staebler05} where the eigenfunctions do not include the asymmetrization due to $\gamma_E$.

The ITG dispersion relation $\frac{\omega}{\omega-\omepi}=-\frac{2n\bar\omega_d}{\omega-\omen}$ was used in (\ref{eq:shift}) to ensure that the shift stays small according to the assumption that the turbulence is ballooned around $\theta=0$ in the same spirit as what is done in \citep{garb02}. It is otherwise determined to cancel to linear terms in $x$ in (\ref{eq:lim_fluide2})
As $\x_0$ is complex, an imaginary shift in $x$ corresponds to a real shift in $k_r$ which means a \emph{linear} stabilization of large radial structures. Strong dependencies of the Gaussian shift are on:
\begin{itemize}
\item $\exb$ shear through the ``$\gamma_E^N$'' term;
\item the parallel velocity gradient through the ``$n\omeu$'' term;
\item the parallel velocity through the ``$\upar$'' term.
\end{itemize}
These dependencies are detailed and compared to \GKW{} self-consistent solutions in the next section. 

\subsection{Linear eigenfunctions validation}
\label{subsec:comp eigenfunctions}
Now that the model employed to predict the linear eigenfunctions has been described, it remains to be compared to self-consistent gyrokinetic eigenfunctions. This comparison is realized with the linear version of the \GKW{} code \citep{peetCPC09} which uses a $\delta f$ decomposition of the distribution function like \QLK. Field aligned coordinates \citep{hamada59} are employed rather than the ballooning representation. There are no approximation in the integration over the pitch-angle and the energy and various magnetic equilibria are available in \GKW. For consistency with \QLK, all direct comparisons are realized with the $\hat{s}-\alpha$ equilibrium in \GKW{} using $\alpha=0$. In this equilibrium, \GKW{} parallel coordinate $s$ is equivalent to \QLK{} $\frac{\theta}{2\pi}$\citep{peetCPC09}. The effects of the parallel velocity and its gradient are shown to be correctly accounted for in \QLK. The effect of $\gamma_E$ is studied as well.

\begin{figure}%
\centering
		\includegraphics[width=0.5\textwidth]{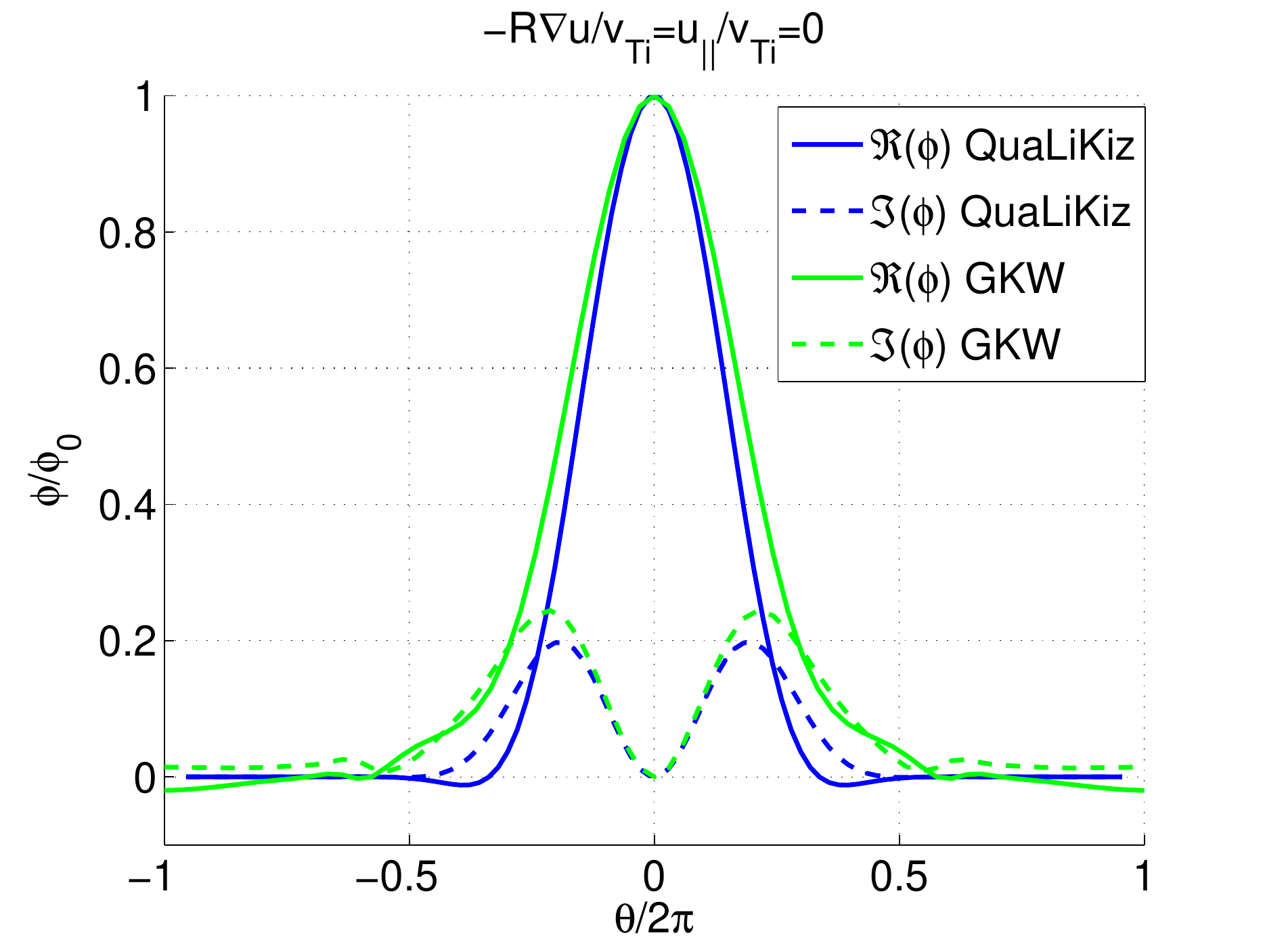}
		\caption{Parallel structure of the eigenfunctions showing null $\kpar$ at zero rotation. GA-std parameters, $\kthe\rho_s=0.3$\label{fig:comp_phi_At9}}
\end{figure}

First, it is verified in Figure~\ref{fig:comp_phi_At9} that the new model previously presented gives a satisfactory agreement with gyrokinetic eigenfunctions in the absence of rotation as in \citep{citrin12,roma07}. Both \GKW{} (in light green) and \QLK{} eigenfunctions (in darker blue) are plotted as a function of the parallel label $\theta/(2\pi)$. GA-std parameters are used. Unless stated otherwise $\epsilon=1/6$, $R/L_n=3$, $R/L_T=9$, $q=2$, $\hat{s}=1$, $Z_{\textsl{eff}}=1$. The poloidal wave number for the study is $\kthe\rho_s=0.3$ as it roughly corresponds to the spectral peak of non-linear fluxes. Figure~\ref{fig:comp_phi_At9} shows a good match between \QLK{} trial eigenfunctions and \GKW{}. \QLK{} eigenfunction is more peaked around $\theta=0$ traducing a slight overestimation of the mode width. This is consistent with Figure~16 from \citep{citrin12}.

\begin{figure*}%
		\subfigure{\includegraphics[width=0.5\textwidth]{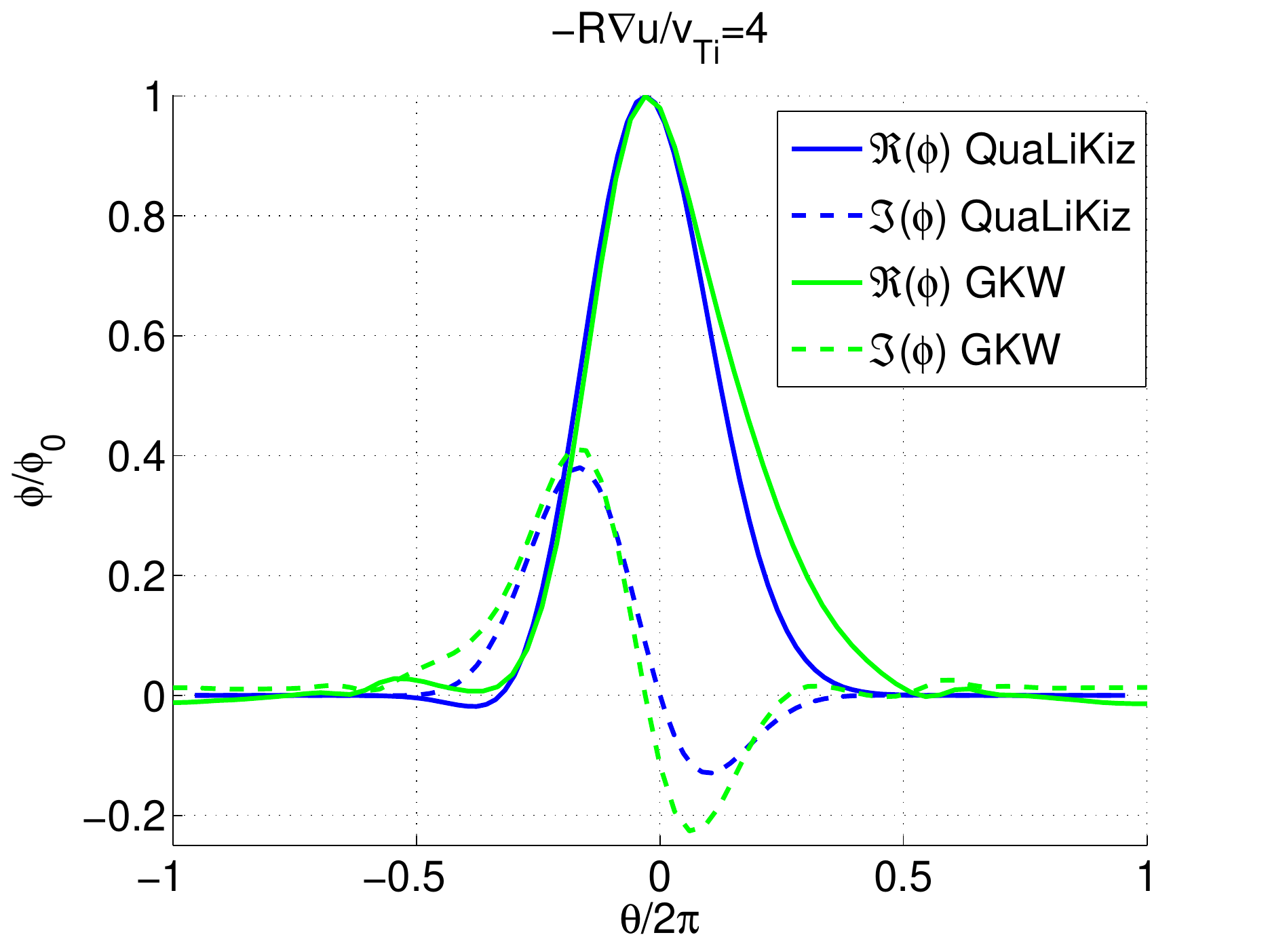}}
		\subfigure{\includegraphics[width=0.5\textwidth]{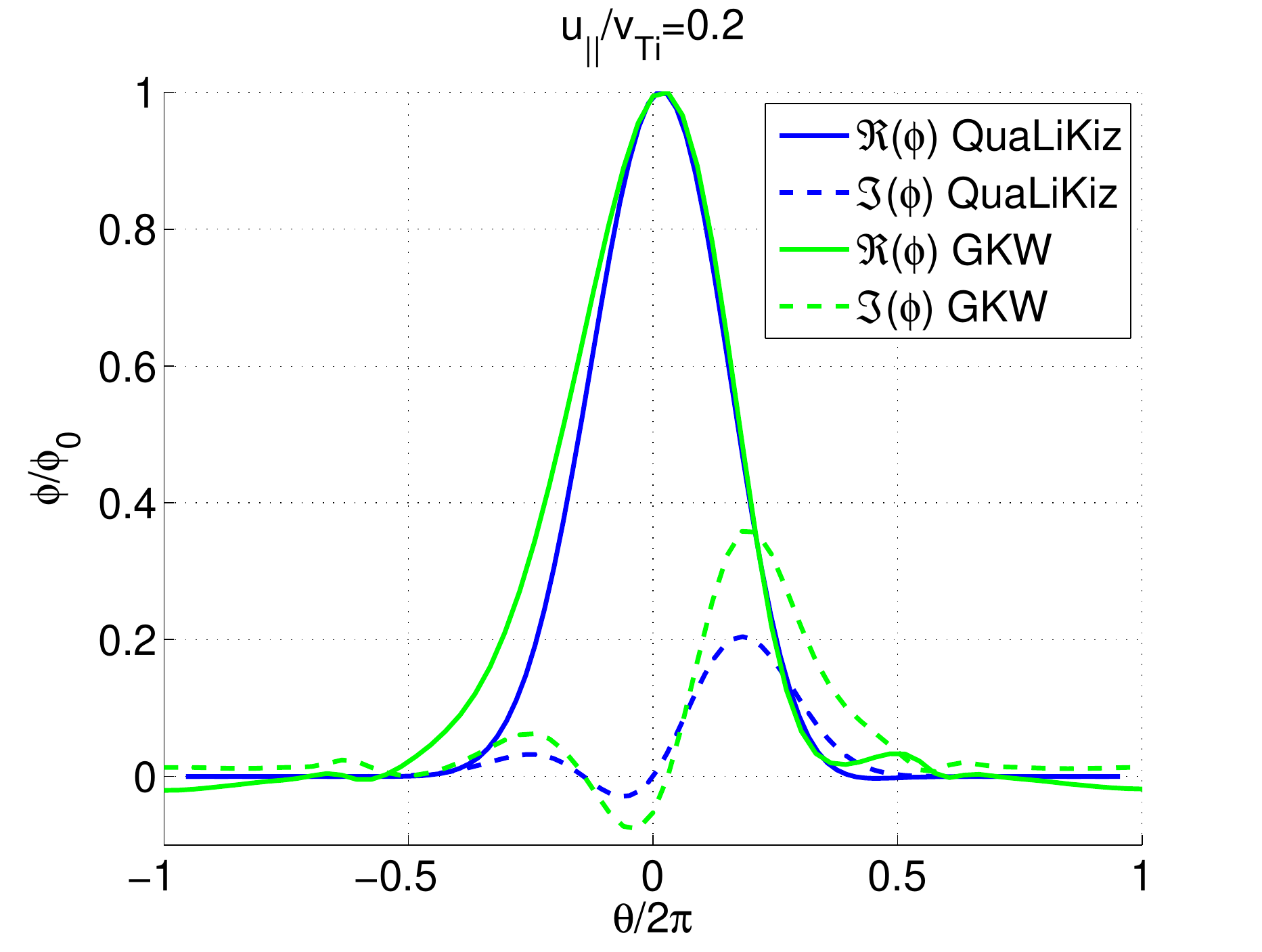}}
		\caption{\label{fig:comp_phi_scan_Au_u_At9}Parallel structure of the eigenfunctions showing finite $\kpar$ in presence of finite $\gradu$ (left) and $\upar$ (right)}
\end{figure*}

\begin{figure}%
\centering
		\includegraphics[width=0.5\textwidth]{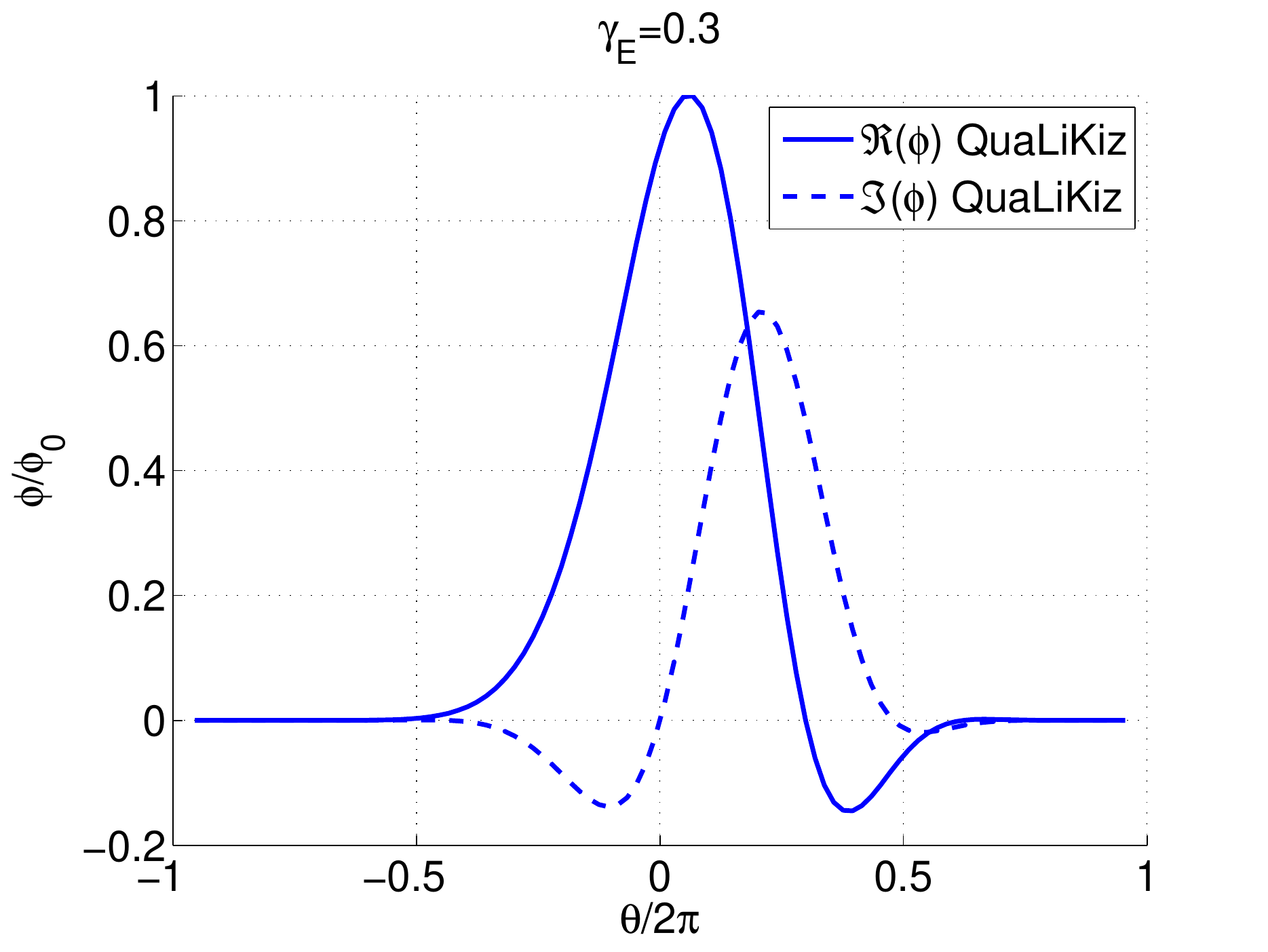}
		\caption{\label{fig:comp_phi_scan_gamE_At9}Parallel structure of the eigenfunctions showing finite $\kpar$ in presence of finite $\exb$ shear}
\end{figure}

The influence of the parallel rotation on the parallel structure of the eigenmodes is now studied in Figure~\ref{fig:comp_phi_scan_Au_u_At9}. In the left panel, \QLK{} and \GKW{} eigenfunctions are plotted against $s=\theta/2\pi$ with GA-std parameters except the parallel velocity gradient (PVG) set to $-4v_{Ti}/R$. This corresponds to maximum experimental values of PVG in core tokamak plasmas\citep{devries08,peet05}. 
In the right panel, the PVG is null and the parallel velocity is set to $0.2 v_{Ti}$. It corresponds to the standard rotation of core plasmas. In both panels, the eigenfunctions appear ballooned in the region where $\theta\sim 0$ confirming previous approximations. But, contrary to the case where there is no rotation \citep{citrin12} (see Figure~\ref{fig:comp_phi_At9}), the eigenfunctions are no longer $\theta$-symmetric. As expected from the expression (\ref{eq:shift}) for the mode shift, $\x_0$ is proportional to $\upar$, $\gradu$ and $\gamma_E$. The agreement with gyrokinetic eigenfunctions is very good in these conditions for both the real and the imaginary parts. The existence of an imaginary part is a novelty. It was previously neglected since, in the absence of sheared flows, the imaginary part of the mode width is small compared its real part and there is no shift in this case (see Figure~\ref{fig:comp_phi_At9}). It was included here because it becomes of the order of the real part in case of strong $\exb$ flow shear. An example of the eigenfunctions found in presence of $\exb$ shearing is plotted in Figure~\ref{fig:comp_phi_scan_gamE_At9} where the imaginary part $\Im(\phi)$ (dashed curve) is found to be comparable to the real part $\Re(\phi)$ of the eigenfunction. The $\theta$-shift of the real part of $\tilde\phi$ is especially important because it represents a $\kpar$-shift contributing to the momentum flux as shown in Sec.~\ref{sec:fluxes}. 
For $\exb$ shear, there is no direct comparison possible, since the general solutions of the linearized gyrokinetic equation in such conditions are oscillating Floquet modes\citep{cooper88,waltz98}. In the reduced model presented here, eigenfunctions are found thanks to the truncation at lowest order of the ballooning representation. 

With the GA-std case set of parameters, chosen for the cases presented above, Ion Temperature Gradient (ITG) modes are dominant. They are known to be ballooned around $\theta=0$ in ballooning space \citep{candy04} so the approximations taken in Sec.~\ref{sec:gyrok} is correct. The case of Trapped Electron eigenmodes (TEM) is briefly discussed now and in more detail in \ref{app:TEM}. TEM are more extended in $\theta$ than ITG modes\citep{brunner98}. Taking only the lowest term of the ballooning representation as is done in \QLK{}, fails to reproduce modes presenting an extension in ballooning space larger than $\theta\in [-\pi; \pi]$ which is especially the case for strongly dominant TEM at $\kthe\rho_s\sim1$. This leads to the overestimation of the TEM stability in this spectral range as illustrated in Figure~\ref{fig:comp_GKW_scanu_At9} by \QLK{} underestimation of the growth rates compared to \GKW{}. For transport studies however, the low $\kthe\rho_s$ matter most and the quasi-linear approximation is only valid at low $\kthe\rho_s$ \citep[see][]{citrin12}. This induces that \QLK{} is able to model correctly TEM dominated regimes as illustrated by Figure~9 of \cite{casati09}.

To summarize, the effects of $\upar$, $\gradu$ and $\exb$ shear are included in the model presented in Sec.~\ref{subsec:fluid model}. They result in a complex shift of the Gaussian eigenfunction and an increase of the relative amplitude of its imaginary part. The influence of $\upar$ and $\gradu$ is successfully benchmarked against \GKW{}. \QLK{} model represents correctly ITG dominated eigenmodes but it cannot capture the extension outside $|\theta|=\pi$ of TEM. This is a necessary trade off to gain two orders of magnitude in CPU time with respect to self-consistent gyrokinetic eigenfunctions calculation making \QLK{} suitable for integrated modeling. 
%\begin{figure}
%	\centering
%		\includegraphics[width=0.5\textwidth]{comp_phi_GKW_TEM_An0_k02.pdf}\hfill
%		\includegraphics[width=0.5\textwidth]{comp_phi_GKW_TEM_An0_k1.pdf}
%		\label{fig:comp_phi_TEM_An0}
%		\caption{Parallel structure of the eigenfunctions showing the increased $\theta$ spreading with $\kthe\rho_s$ in the case of TEM. $R/L_{Ti}=0$, $R/L_n=0$ and other parameters from GA-std test case. $\kthe\rho_s=0.2$ left panel. $\kthe\rho_s=1.0$ right panel.}
%\end{figure}
\section{Impact of sheared flows on linear growth rates}
\label{sec:growth rates}
%In Sec.~\ref{sec:fluid}, the fluid model for the eigenfunction calculation has been presented and its results have been validated against the gyrokinetic code \GKW. Additional simplifications are performed in \QLK{} from the linear gyrokinetic theory as detailed in Sec.~\ref{sec:gyrok}. The low Mach number approximation is taken, as well as the lowest order ballooning representation to reduce the size of the problem, a large aspect ratio is considered and the $\hat{s}-\alpha$ magnetic equilibrium is employed. Moreover trapped and passing particles are treated separately and the Bessel functions from the gyrokinetic transformation is integrated separately in energy. 
A way to validate the model developed in Sec.~\ref{sec:gyrok} and \ref{sec:fluid} is to compare the linear growth rates $\gamma=\Im(\omega)$ found with \QLK{} against the results from a gyrokinetic code which does not use the simplifications previously detailed. An important benchmark effort has already been done, comparing \QLK{} growth rates against \gst{} \citep{bour02,roma07} and \Gene\citep{citrin12}. The comparison is limited here to the sheared flows impact by varying $\upar$, $\gradu$ and $\gamma_E$ using \GKW{} linear simulations and GA-standard based test cases. Unless stated otherwise $\epsilon=1/6$, $R/L_T=9$, $R/L_n=3$, $q=2$, $\hat{s}=1$, $\alpha=0$, $\nu^\ast=0$ in this section. The parallel velocity gradient destabilization and the stabilizing effect of $\exb$ shear are successfully benchmarked. The effects of the parallel velocity are recovered within the range of validity of the low Mach number approximation.

\subsection{Parallel velocity gradient instability with $\gradu$}
\label{subsec:PVG}
\begin{figure}%
\centering
\includegraphics[width=0.5\textwidth]{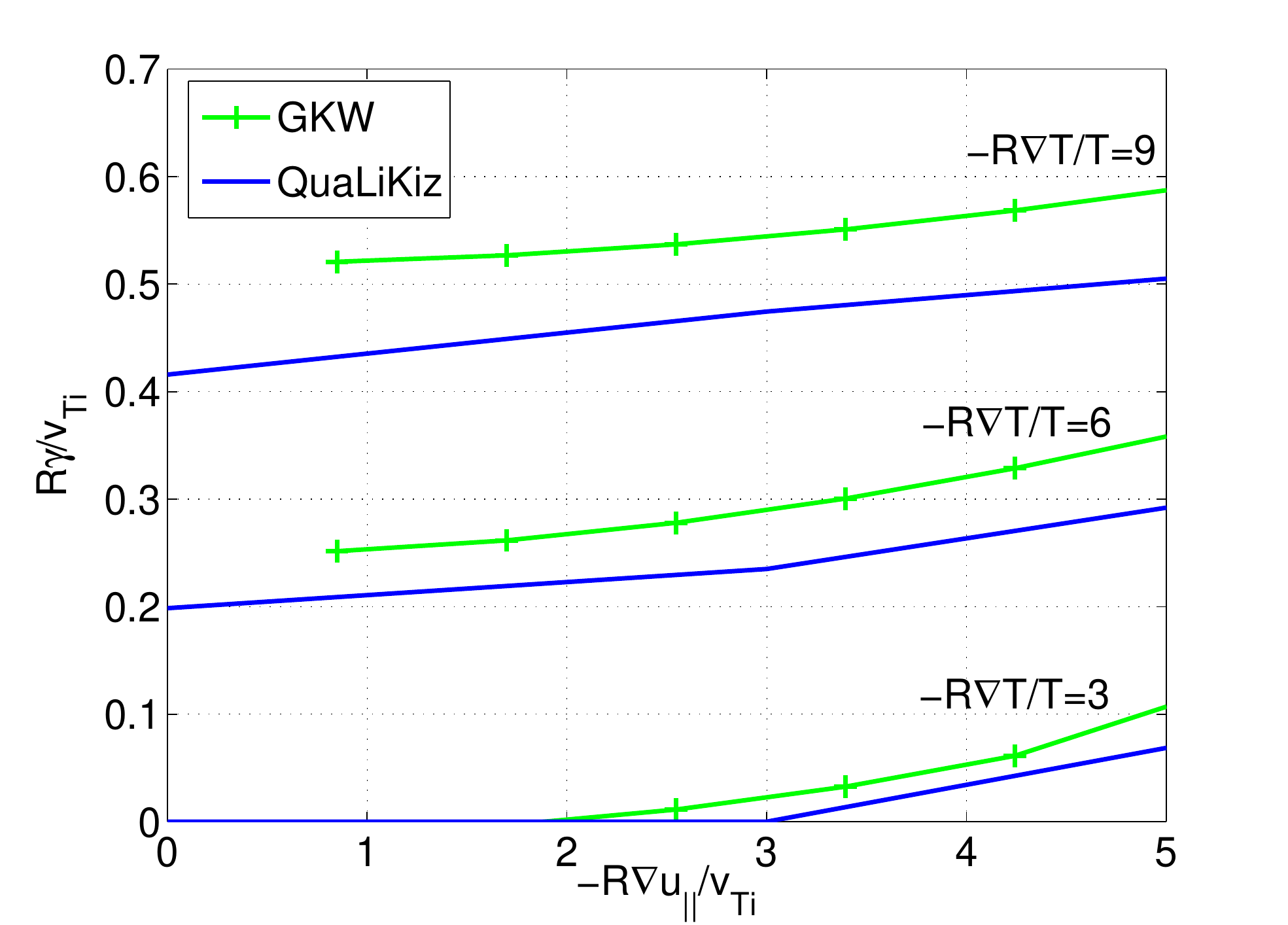}
	\caption{\label{fig:comp_GKW_scanAu_At9}Maximum linear growth rates from \QLK{} and \GKW{} for GA-std parameters}
\end{figure}

First, let us concentrate on $\gradu$. It has been extensively reported in the literature that parallel velocity gradients (PVG) destabilize a Kelvin-Helmholtz like instability\citep{dangelo65,garb02,peet05}. PVG instabilities are destabilized by velocity gradients at rather high values $\gradun\approx 5$ compared to the experiments\citep{peet05}. But its threshold is reduced with increasing temperature gradient so that it can destabilize otherwise marginally stable conditions for ITG turbulence. Finally, PVG is known for enhancing the growth rates of already unstable ITG modes. All these effects are presented in Figure~\ref{fig:comp_GKW_scanAu_At9} where a scan in $\gradu$ is performed up to $\gradu=-5v_{Ti}/R$ for 3 values of temperature gradients $R/L_T=\{3, 6, 9\}$. For flatter temperature profile conditions ($R/L_T=3$), which is linearly stable without rotation, the PVG destabilization threshold is recovered. For the peaked temperature profile condition ($R/L_T=\{6,9\}$), which are ITG unstable without rotation, the growth rate inflation with $\gradu$ is captured by \QLK. The values of the growth rates are nevertheless slightly underestimated.

\subsection{Impact of $\upar$}
\label{subsec:mach}
\begin{figure*}%
\centering
\subfigure{\includegraphics[width=0.5\textwidth]{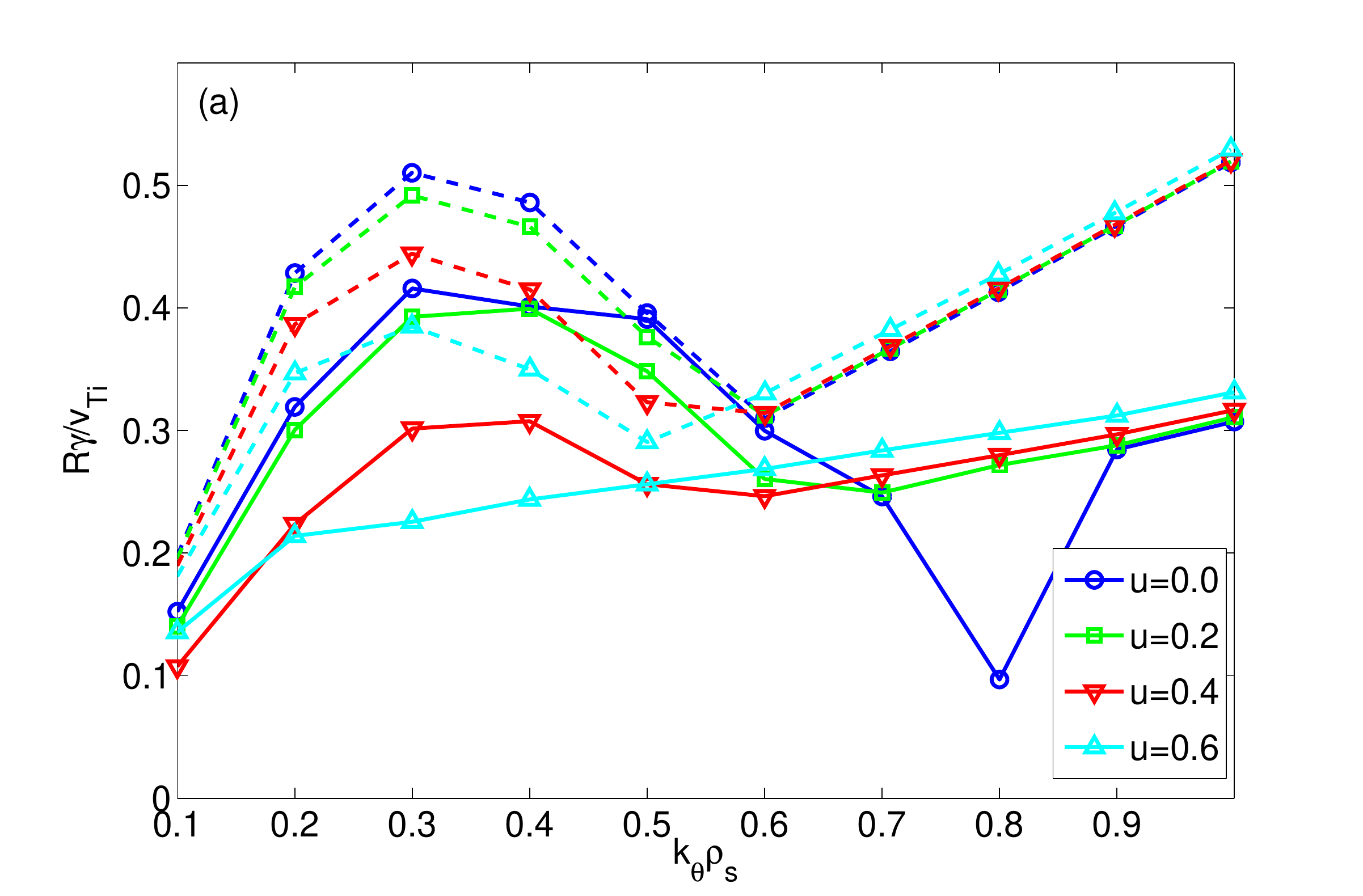}\label{fig:comp_GKW_scanu_At9_nocent}}%\hfill
\subfigure{\includegraphics[width=0.5\textwidth]{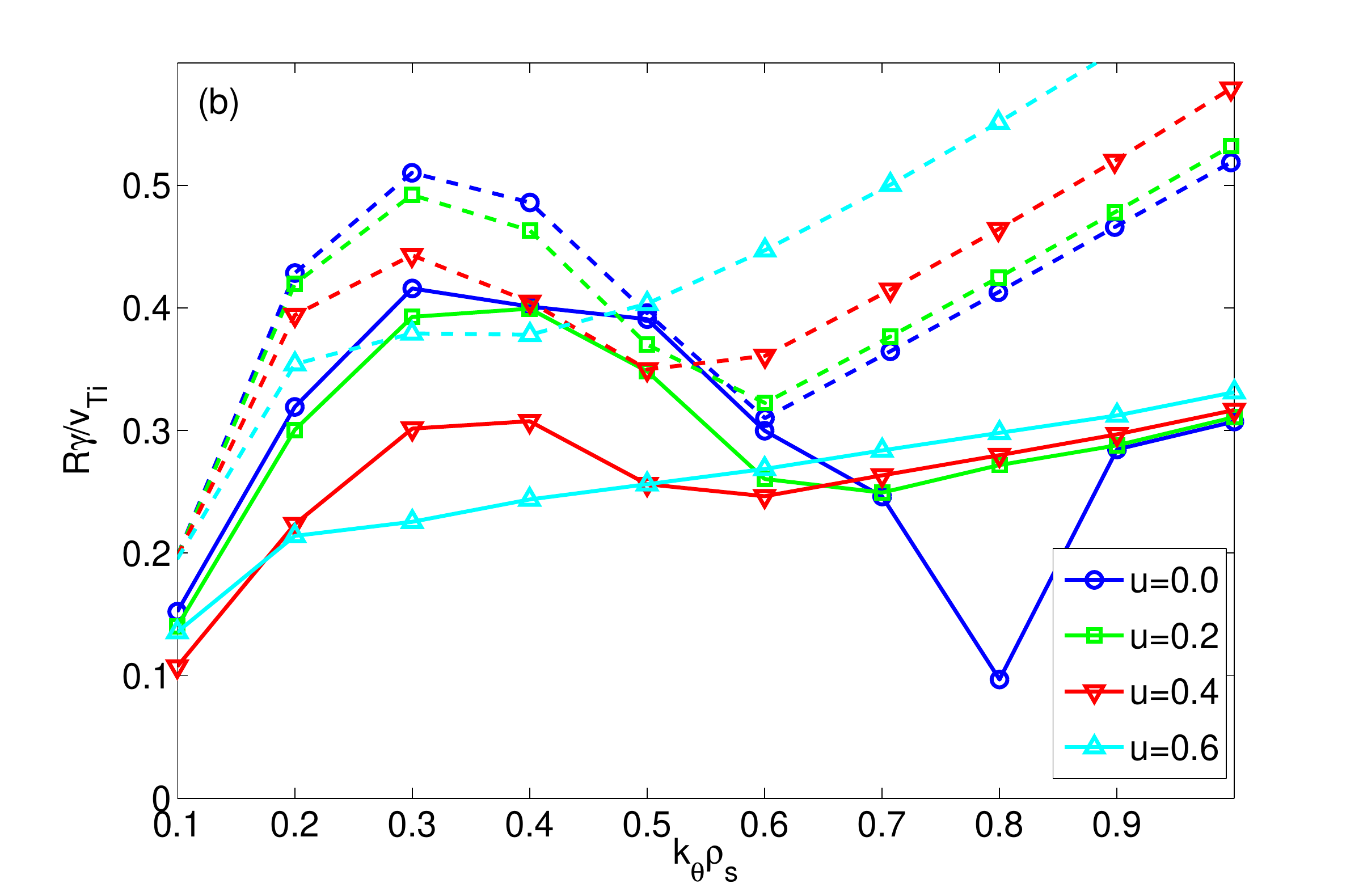}\label{fig:comp_GKW_scanu_At9_cent}}
	\caption{\label{fig:comp_GKW_scanu_At9}Linear growth rates from \QLK{} (plain curves) and \GKW{} (dashed) for GA-std based cases with various $u=\uparn$ values. (a) \GKW{} run without centrifugal effects (b) \GKW{} run with centrifugal effects}
\end{figure*}

The parallel velocity is known to have opposite effects on ions and electrons modes. It stabilizes ITG modes and destabilizes trapped electron modes (TEM) via the expansion of the trapped domain in velocity space with increasing $\upar$\citep{cass10,erratum_cass10}. These effects are studied in Figure~\ref{fig:comp_GKW_scanu_At9}. Simulations from \QLK{} (in plain curve) and \GKW{} (in dashed curve) based on GA-std parameters are represented. The parallel velocity is varied from $0$ to $0.6v_{Ti}$, a larger value than usually observed in high aspect ratio tokamak core plasmas\citep{devries08}. The effect of the low Mach number approximation -- used in \QLK{}, not in \GKW{} -- is analyzed. 

When comparing \GKW{} (with centrifugal effects) and \QLK{}, Figure~\ref{fig:comp_GKW_scanu_At9_cent}, it is clear that ITGs are stabilized in both codes but TEMs are not destabilized in \QLK. This discrepancy is due to the low Mach number approximation which does not retain centrifugal effects. They were removed in \GKW{} in Figure~\ref{fig:comp_GKW_scanu_At9_nocent} to illustrate this. Indeed, without centrifugal effects, \GKW{} electron modes are not destabilized. Moreover, at higher $\uparn$ values, ITGs are overstabilized in \QLK{} and TEMs become dominant for lower $\kthe\rho_s$ values as $\upar$ increases due to the stabilization of ITGs. This is a consequence of the development up to second order in $\upar$ of the equilibrium distribution function (see Equation \ref{eq:electroneutralite}) which underestimates the values of the exponential in $\upar$ contained in $f_0$ definition at larger values of $\vpar$. The underestimation of TEM growth rates by \QLK{} at higher $k_\theta\rho_s$ for any values of $\upar$ is related to a discrepancy between \QLK{} and \GKW{} eigenfunctions as detailed in \ref{app:TEM}. 

\subsection{Stabilization by $\exb$ shear}
\label{subsec:ExB}
\begin{figure}
\centering
	\includegraphics[width=0.5\textwidth]{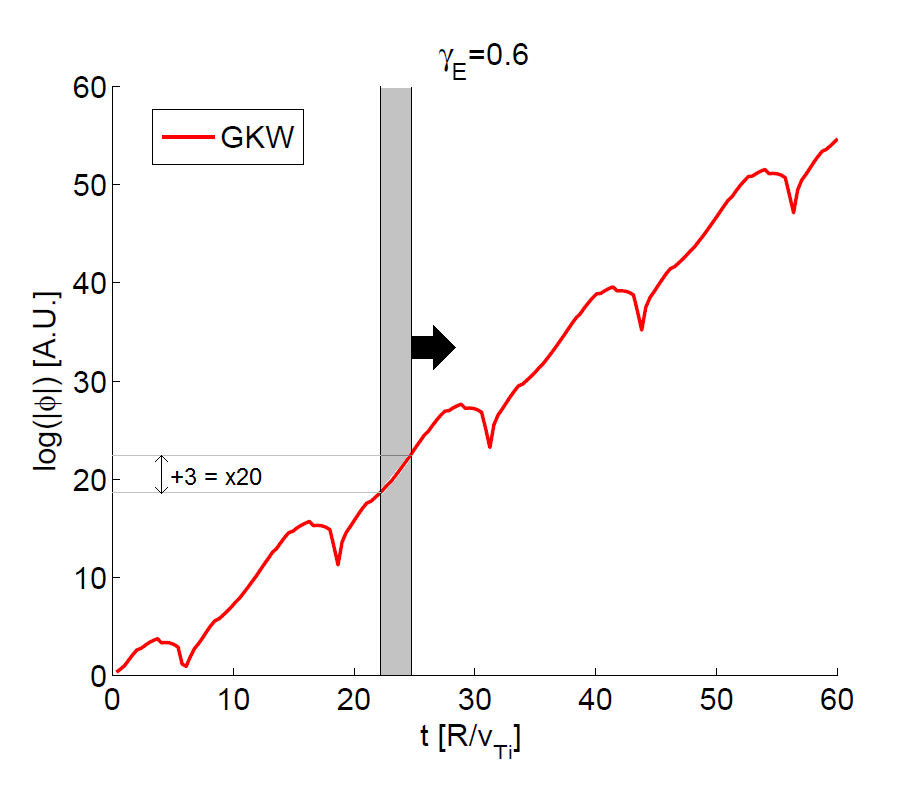}
	\caption{\label{fig:floquet}Example of the time evolution of a Floquet mode from a GKW simulation at $R/L_T=15$, $\gamma_E=0.6$ and other parameters from GA-std. The shaded region corresponds to $3\gamma^{-1}\approx3\tau_{NL}$. The black arrow represents the displacement of the shaded region along $t$.}
\end{figure}

\begin{figure}
\centering
	\includegraphics[width=0.5\textwidth]{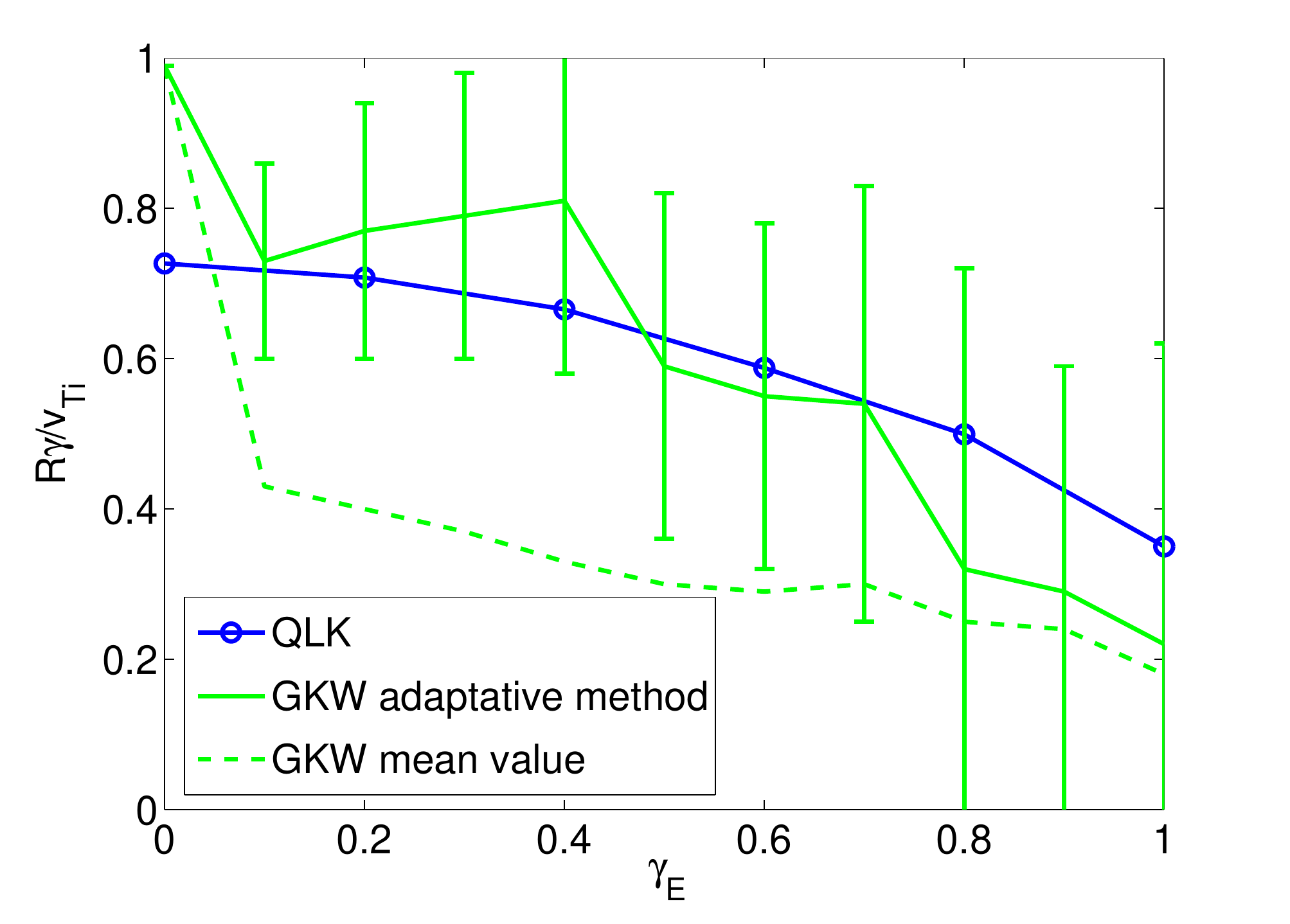}
	\caption{\label{fig:comp_fluide_GKW_scanvEAt15}Maximum \QLK{} growth rates and \GKW{} effective growth rates calculated with the standard averaging method and a new statistical method}
\end{figure}

The extensively studied stabilization of the turbulence by $E\times B$ shear \citep{biglari90,dong93,waltz94,hahm95,waltz98,roach09,barnes11} is addressed in this section. To be able to perform the comparison with \GKW{}, we highlight that a new method to calculate effective growth rates for initial value codes such as \GKW{} with $\exb$ shear is developed. This method is close to that of \citep{citrin13} and results in a better qualitative agreement with non-linear observations. Indeed, with finite $\exb$ shear, Floquet modes are solutions of the linearized gyrokinetic equation, composed of an exponentially growing part and an oscillating part. Consequently, when averaging over the entire temporal window of the simulation, a strong drop in the effective growth rate is observed for the first non-zero value of $\gamma_E$ and then a weak dependence with $\gamma_E$ is seen as explained in \citep{waltz98} and represented in Figure~\ref{fig:comp_fluide_GKW_scanvEAt15} dashed curve. In contrast, non-linear simulations show a smooth reduction of the fluxes with increasing $\exb$ shear\citep{cass09}, fitted at times by a linear quench rule\citep{waltz98}. An explanation for this discrepancy is that the non-linear decorrelation time is shorter than the time over which one averages the growth rates. The method proposed here to resolve this issue can be decomposed in two steps illustrated by Figure~\ref{fig:floquet}.
\begin{itemize} 
\item First, an effective growth rate $\gameff(t)$ is calculated on 3 decorrelation times $\tau_{NL}$ considering that $\tau_{NL}=\gameff^{-1}$. It means that $\gameff=\left(\ln(\phi(t+\Delta t))-\ln(\phi(t))\right)/\Delta t$ is calculated with $\Delta t=3/\gameff$. Equivalently $\phi(t+\Delta t)=\exp(3)\phi(t)$. The corresponding $\Delta t$ is represented by the shaded area in Figure~\ref{fig:floquet}; 
\item The time window corresponding to $3\tau_{NL}$ is then moved along the simulation as indicated by the black arrow in Figure~\ref{fig:floquet}. The effective growth rate of the entire simulation is taken to be the $3^{rd}$ quartile of the ensemble of $\gameff[0;t_{end}]$ to remove all the negative $\gameff(t)$ from the statistics.
\end{itemize}
This method is compared to the standard one -- see for example \citep{roach09} -- in Figure~\ref{fig:comp_fluide_GKW_scanvEAt15}. The so-called  ``GKW mean value'' dotted curve represents the usual method and the ``GKW'' plain curve with error bars represents the method described above. The error bars extent corresponds to one standard deviation around the  $3^{rd}$ quartile value. The usual ``jump'' in $\gamma$ from 0 to finite value of $\gamma_E$ is reduced, resulting in better qualitative agreement with the results from non-linear simulations. The growth rates from the eigenvalue code \QLK{} are plotted on the same figure in plain curve for comparison. They are in agreement with $\gameff$ within the error bars of the method presented above. This result shows that the $\exb$ stabilization mechanism is captured by \QLK{} approach using fluid shifted Gaussian eigenfunctions without any fitting parameter contrary to the quench rule usually used in transport codes \citep{waltz97,staebler05}.

Through the three examples presented above, \QLK{} linear growth rates evolution with the three relevant quantities for sheared flows in a tokamak plasmas -- $\upar$, $\gradu$ and $\gamma_E$ -- have been validated. Along with the correct linear eigenfunctions, this gives the possibility to make a quasi linear estimate of the turbulent heat, particle and momentum fluxes accounting properly for PVG and $\exb$ shear stabilization at lower CPU cost.

\section{Quasi-linear fluxes}
\label{sec:fluxes}
Quasi-linear models are extensively used to predict heat, particle and momentum fluxes without the numerical cost of non-linear simulations\citep{angioni04,angioni12,camenen09, casati09,dannert05,kinsey05b,kinsey08}. They have been heavily benchmarked against non linear simulations for heat and particles \citep{merz10,staebler07,waltz_casa09} and more recently for momentum\citep{staeb13}. In this section the quasi-linear momentum flux is derived in \QLK{} formalism. In \ref{subsec:mom_flux_exp} the linear response is shown to be similar to the expressions of $\mathcal{L}_{s,\text{pass}}$ and $\mathcal{L}_{s,\text{tr}}$ of the linearized gyrokinetic equation (\ref{eq:vlasov}). Indeed, in the quasi-linear approximation, the fluxes can be written as derived in App. A of \citep{bour07}:
\begin{equation}
\label{eq:flux}
\Gamma=\sum_{\n,\omega}\n\cdot\Im\left(\frac{\n\cdot\partial_\J f_0}{\omega-\OmeJ+\imath o^+}\right)|\tilde{h}_{\n\omega}|^2
\end{equation}
$|\tilde{h}_{\n\omega}|^2$ corresponds here to the saturated potential. This potential cannot be self-consistently determined since there is no saturation mechanism embedded in the theory. It must be constructed based on experimental observations and non-linear simulations\citep{bour07,casati09}. In \QLK{}, the saturated potential maximum is defined by a mixing length rule discussed in Sec.~\ref{subsec:saturated potential}. The saturated potential spectrum in $k_\bot$ is also reviewed in Sec.~\ref{subsec:saturated potential}. The results are compared against non-linear \GKW{} simulations in \ref{subsec:comp flux}. The $\exb$ shear quenching of the particle and heat fluxes is recovered. The associated momentum fluxes match for small values of $\exb$ shear $\gamma_E^N<0.1$ but overestimated in \QLK{} by a factor 2 for larger values of $\gamma_E$. Finally the influence of $\upar$ and $\gradu$ on \QLK{} momentum flux is validated by calculating the Prandtl and pinch numbers.
\subsection{Quasi-linear momentum flux in \QLK{} formalism}
\label{subsec:mom_flux_exp}
As indicated in (\ref {eq:flux}), quasi-linear fluxes are composed of two parts. One is a linear response and the other is the saturated potential. The linear response is detailed here. In an axisymmetric tokamak, the flux surface averaged \emph{toroidal} momentum flux is the quantity to calculate since the flux surface averaged angular momentum $p_\phi=\int{mRv_\phi\tilde{f}\d^3v}$ is globally conserved\cite{abiteboul11}. Here its perpendicular part is neglected and only the \emph{parallel} contribution is retained. Moreover, $R$ used in the definition of the momentum flux $\Pi_\parallel$ is the major radius at the magnetic axis. There is therefore no $\epsilon$ correction of this quantity. The momentum flux calculated in \QLK{} is defined as follows:
\begin{equation}
\Pi_\parallel=\sum_s\Re\left\langle m_sR\vpar\tilde{f}_{s}\frac{\imath k_\theta\tilde\phi}{B}\right\rangle
\end{equation}
$\tilde{f}_s=\frac{\n\partial_\J f_0^s}{\omega-\n\OmeJ+\imath o^+}\tilde{h}$ is the perturbed distribution function determined by the linearized Vlasov equation and $\langle\cdots\rangle$ means integration over the velocity space. $\Pi_\parallel$ is positive for an outward flux of momentum in the direction of $\mathbf{B}$. Using the formalism developed in Sec.~\ref{sec:gyrok}, the complete expression of $\Pi_\parallel$ is presented in (\ref{eq:mom_flux}) by replacing $\tilde{f}_s$ with its expression given in \ref{app:mom_flux} (\ref{eq:electroneutralite}).

Apart from the saturated potential $\tilde\phi_{n\omega}$, the expression (\ref{eq:electroneutralite}) is similar to the linear gyrokinetic response presented in Sec.~\ref{sec:gyrok} except that only the imaginary part is of interest for the flux and that the integrations over ($\xi$, $\lambda$) are slightly different due to the multiplication by $v_\parallel=\pm v_{Ts}\sqrt{\xi(1-\lambda b)}$. The same techniques as before are then employed. The contributions from trapped and passing particles to the momentum flux are treated separately. The expression for $\mathcal{J}_{s,pass}$ is detailed in (\ref{eq:mom_flux_pass}). Note that the parity of (\ref{eq:mom_flux_pass}) is opposite to that of (\ref{eq:passing}) due to the multiplication by $v_\parallel$. This guarantees that without rotation the momentum is zero. For trapped particles, the multiplication by $v_\parallel$ implies there is no contribution to the momentum flux at lowest order in $\epsilon$. However, when expanding up to first order in $\sqrt{\epsilon}$, there is a net contribution from trapped particles, detailed in (\ref{eq:mom_flux_tr}).

Given the expressions of the passing and trapped particle contributions to the momentum flux, (\ref{eq:mom_flux2}) can formally be written in the form:
\begin{equation}\label{eq:mom_flux_detail}
\Pi_\parallel=\sum_sm_sn_sR(-\chi_\parallel\gradu+V_\parallel\upar)+\Pi_{RS}
\end{equation}
$\chi_\parallel$ representing the momentum diffusivity, $v_\parallel$, the momentum pinch and $\Pi_{RS}$ being the residual stress.
However, the identification of $\chi_\parallel$, $V_\parallel$ and $\Pi_{RS}$ with (\ref{eq:mom_flux}) is not as straightforward as it may appear. From (\ref{eq:mom_flux_pass}) and (\ref{eq:mom_flux_tr}), it is clear that $\Pi_\parallel$ contains terms directly proportional to $\upar$ and $\gradu$. They are called $\Pi_u$ and $\Pi_{\nabla{u}}$. They do not contain all contributions from $\upar$ and $\gradu$. The remaining terms are proportional to the linear eigenfunction shift $\x_0$ which, itself, is proportional to $\gradu$, $\upar$ and $\gamma_E$ as expressed by (\ref{eq:shift}) from Sec.~\ref{sec:fluid}\citep{gurcan07}. These terms proportional to the eigenfunction shift are called $\Pi_{\mathrm{x0}}$. If $\exb$ shear is the only symmetry breaker, $\Pi_{\mathrm{x0}}\equiv\Pi_{RS}$. Otherwise, $\Pi_{\mathrm{x0}}\propto \upar, \gradu, \gamma_E$ cannot be identify with $\Pi_{RS}$ as $\Pi_{\nabla{u}}$ (resp. $\Pi_u$) does not contain all conductive (resp. convective) contributions to the momentum flux. 

The different contributions can be separated by linear regressions. Since we are searching for three unknowns, three simulations are performed with the same set of parameters except for $\upar$, $\gradu$ and $\gamma_E$. The first one is the test simulation. The second one is performed with the parallel velocity modified by $\pm 20\%$. Both the parallel velocity gradient and the $\exb$ shearing are affected by this modification of the parallel velocity. The last simulation is performed with the parallel velocity \emph{incremented} by $\pm0.05v_{Ti}$. The parallel velocity gradient is not perturbed by this modification. Considering that such modifications have a \emph{linear} effect on the momentum flux, a linear regression is possible to estimate the momentum diffusivity $\chi_\parallel$, the pinch $V_\parallel$ and the residual stress $\Pi_{RS}$. If $\Pi_1$ is the parallel momentum flux from the $1^{\text{st}}$ simulation, $\Pi_2$ from the $2^{\text{nd}}$ and $\Pi_3$ from the $3^{\text{rd}}$, and under the assumption that the changes presented above induce only a linear modification, they read:
\begin{subequations}
\label{eq:3-point}
\begin{align}
\Pi_1&=\sum_sm_sn_sR(-\chi_\parallel\gradu+V_\parallel\upar)+\Pi_{RS}\\
\Pi_2&=\sum_sm_sn_sR(-1.2\chi_\parallel\gradu+1.2V_\parallel\upar)+\Pi_{RS}\\
\Pi_{3}&=\sum_sm_sn_sR(-\chi_\parallel\gradu+V_\parallel(\upar+0.05v_{Ti}))+\Pi_{RS}
\end{align}
\end{subequations}
The system \eqref{eq:3-point} is a set of 3 independent equations of 3 variables. Therefore each of the variables $\chi_\parallel$, $V_\parallel$ and $\Pi_{RS}$ is uniquely defined. Varying $\upar$ by $\pm20\%$ and incrementing $\upar$ by $\pm0.05v_{Ti}$ defines 3 different sets of equations. If the momentum flux dependence with respect to $\upar$ and $\gradu$ is linear the 3 systems should give the same results. In the opposite case, the dispersion between the results (inversely) measures the validity of the bilinear regression. The method ensures that linear dependencies of $\gamma_E$ with $\gradu$ and $\upar$ are removed from the residual stress and accounted for in $\chi_\parallel$ and $V_\parallel$ respectively. 

The comparison between the direct separation and the 3-point method gives an estimate of the importance of the eigenfunction contribution to the conductive and convective part of the momentum flux as discussed in detail and evaluated in Sec.~\ref{subsec:comp flux}. Concerning the residual stress, it corresponds to the momentum flux induced by the parallel symmetry breakers other than $\upar$ and $\gradu$. In \QLK{}, only the $\exb$ shearing induced residual stress is calculated. Indeed, the global effects from turbulence intensity gradient \citep{gurcan10c} or profile shearing \citep{waltz11} are not included. They produce a residual stress of the same order as $\exb$ shearing by tilting the ballooned structure of the turbulence around $\theta_0\neq0$ \citep{camenen11}.

\subsection{Saturated potential} 
\label{subsec:saturated potential}
The saturated potential is constructed according to experimental observations and non-linear simulations\citep{mckee01,bour07,casati09}. The frequency spectrum is a Lorentzian of width $\gamma$ as explained in \cite{bour07}. In cases of simulations with large $\exb$ shear, the width is modified. Indeed, if $\gamma_E>\gamma$, the shear rate defines a shorter time scale than the linear growth rate. The following rule is therefore: the width of the Lorentzian is $\text{max}(\gamma(k),\gamma_E)$. This rule would need to be validated by non-linear gyrokinetic simulations. It implies a high resolution diagnostic for the frequency that deals correctly with the implementation of the $\exb$ shear. To our knowledge, such a diagnostic does not exist yet. 

\begin{figure*}%
\subfigure{\includegraphics[width=0.5\textwidth]{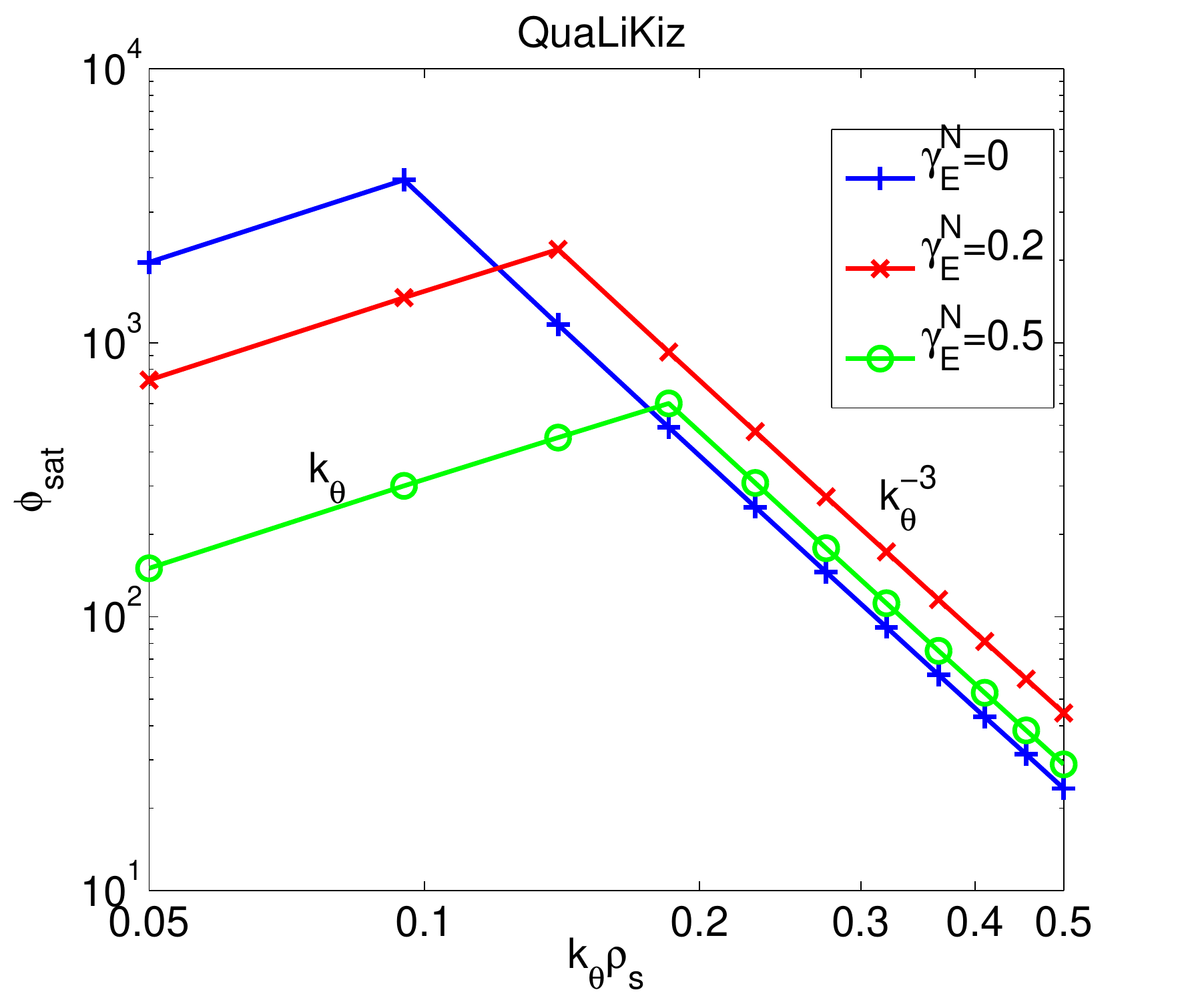}}
\subfigure{\includegraphics[width=0.5\textwidth]{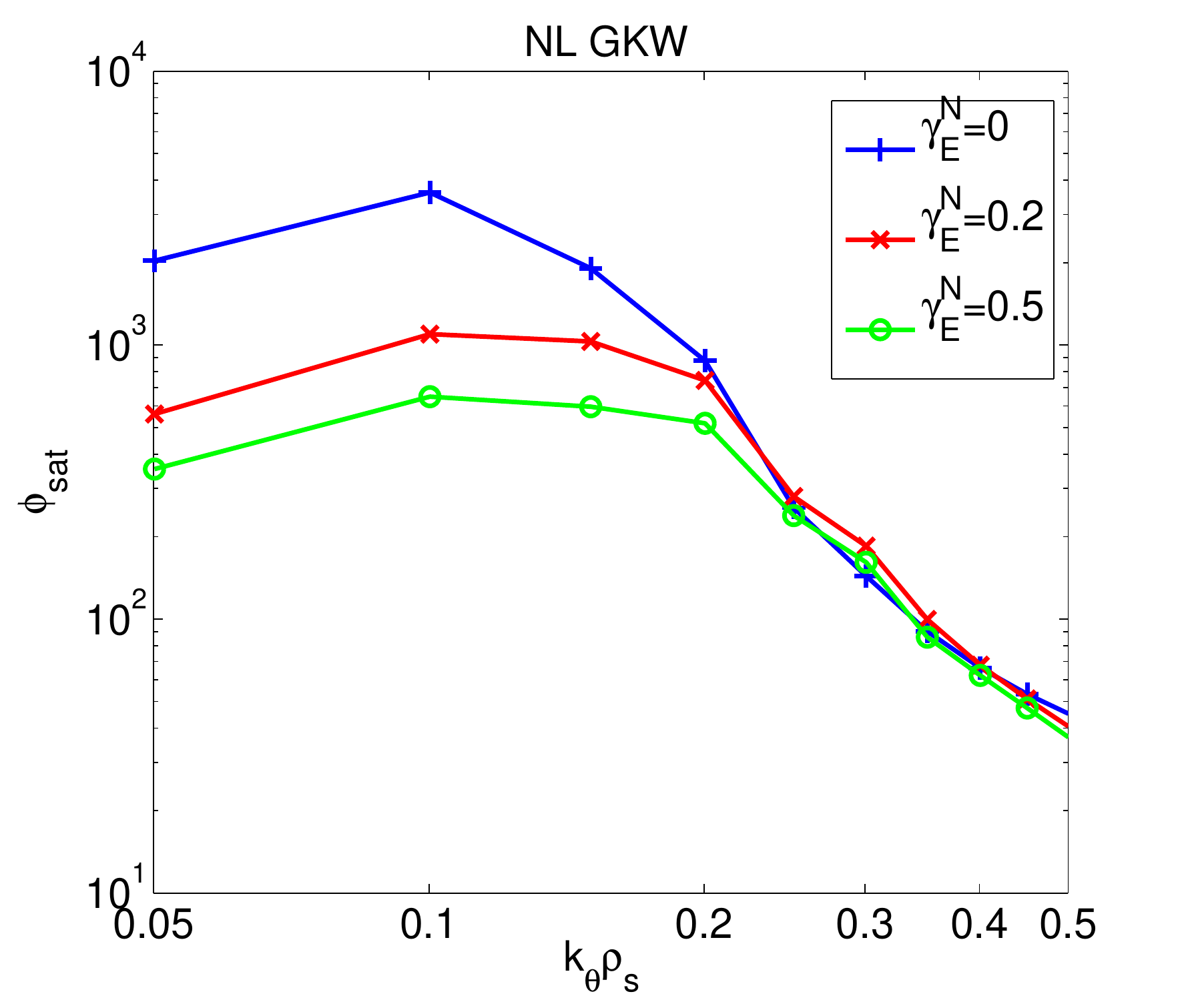}}
\caption{\label{fig:comp_gkw_ky_spec_scangamE}\QLK{} $\tilde\phi_{sat}$ estimate (left panel) and \GKW{} non linear saturated potential. Simulations with $\exb$ shear only. $\gamma_E$ values in $v_{Ti}/R$ units.}
\end{figure*}

For the perpendicular wave number spectrum, it was found that a $k_\bot^{-3}$ spectrum reproduces the cascade towards smaller scales found in non-linear simulations and experimentally measured\citep{casati09}. With such a spectrum, wave numbers higher than $k_\theta\rho_s=1$ will have little influence on the transport level. Indeed, the significant contributions to the turbulent fluxes found in some non-linear simulations at higher wave numbers \citep{jenko02} depart from the estimation of a saturation rule which is used here. Therefore, the wave number range is kept between $k_\theta\rho_s=0.05$ and $k_\theta\rho_s=1$ in the simulations although there is no intrinsic limitation of the maximum perpendicular wave number computable in \QLK{}. For the inverse cascade at larger scales, Figure~\ref{fig:comp_gkw_ky_spec_scangamE} illustrates that a linear spectrum reproduces better non-linear simulations than the $k_\bot^3$ spectrum previously employed. 

It should also be noted that all unstable modes (ITGs and TEM) are taken into account in \QLK{} and not only the dominant mode. The fluxes are made of the sum of all unstable mode contributions. For each unstable mode, a mixing length rule estimate is used to evaluate its quasilinear weight in the fluxes such that there is no free parameters involved. A mixing length rule estimate on the most unstable mode is used to fix the wave number at which the saturated potential is maximum:
\begin{equation}
\text{max}\left(D_{\textsl{eff}}(k_\bot)\approx\frac{R\Gamma_s}{n_s}\right)=\frac{k_\theta e_sR}{B}T_s|\tilde\phi_n|^2\Bigg|_{k_\text{max}}=\text{max}\left(\frac{\gamma}{\langle k_\bot^2\rangle}\right)
\end{equation}
The expression for $\langle k_\bot^2\rangle$, based on the idea proposed in \citep{dannert05}, has been recently revisited in \citep{citrin12} to improve \QLK{} fluxes estimation at low magnetic shear. It reads:
\begin{equation}\label{eq:kperp}
\langle k_\bot^2\rangle=k_\theta^2+k_r^2=k_\theta^2+\left(\sqrt{k_\theta^2\hat{s}^2\langle\theta^2\rangle}+\frac{0.4\exp(-2\hat{s})}{\sqrt{q}}+1.5(k_\theta-0.2/\rho_s)H(k_\theta-0.2/\rho_s)\right)^2
\end{equation}
The expression of $k_r$ in \QLK{} mixing length rule was modified because it was found that, at low magnetic shear, $k_r^2=k_\theta^2\hat{s}^2\langle\theta^2\rangle$ resulting from the magnetic field lines shearing is underestimated with respect to non-linear $k_r$ \citep[see][Sec.~IV C.]{citrin12}. The factor $\frac{0.4\exp(-2\hat{s})}{\sqrt{q}}$ was found to represent best the non linear isotropization at low magnetic shear. Finally, the term $1.5(k_\theta-0.2/\rho_s)H(k_\theta-0.2/\rho_s))^2$ ($H$ is the Heaviside function) is only present for completeness, to ensure the agreement with non-linear simulations at smaller scales which does not participate much to the transport in mixing length models. This definition for the mixing length rule is modified by the linear eigenfunction shift $\x_0$ proportional to the symmetry breakers (\ref{eq:shift}). Indeed, the linear eigenfunction enters the expression of $\langle\theta^2\rangle$ from (\ref{eq:kperp}): 
\begin{equation}\label{eq:theta2}
\langle \theta^2\rangle=\frac{\int\theta^2\tilde\phi\d\theta}{\int\tilde\phi\d\theta}=\frac{2d^2}{\Re(\w^2)}\frac{\Gamma(0.75)}{\Gamma(0.25)}+\frac{\Im(\x_0)^2d^2}{\Re(\w^2)^2}
\end{equation}
Therefore, the symmetry breakers influence $\langle k_\bot^2\rangle$ through the imaginary part of the eigenfunction shift $\Im(\x_0)$ and the real part of the mode width, the latter being proportional to the growth rate found in the fluid model. Thus, both $\gamma$ (see Sec.~\ref{sec:growth rates}) and $\langle k_\bot^2\rangle$ are modified in the presence of finite sheared rotation. This approach is different than that of \citep{staeb13}. Indeed, since there is no parallel asymmetrization of the linear eigenmodes with the $\exb$ shear in TGLF, a non-linear spectral shift model was built to compute the induced momentum flux. Here, the parallel asymmetrization of the linear eigenmodes with the $\exb$ shear fulfills this task, avoiding using a non-linear spectral shift fitting model.

The modification of $\tilde\phi_\textsl{sat}$ induced by $\exb$ shear are plotted and compared to non-linear \GKW{} saturated potential \citep{cass09} in Figure~\ref{fig:comp_gkw_ky_spec_scangamE}. In the simulations presented here, GA-std case parameter set has been employed with $\upar=\gradu=0$. Three values of $\exb$ shear are chosen corresponding to an experimentally relevant range of $\gamma_E$ from 0 to $0.5R/v_{Ti}$. The $\kthe\rho_s$ extent covered in Figure~\ref{fig:comp_gkw_ky_spec_scangamE} corresponds to the transport relevant spectral range.

For both \QLK{} and \GKW, as $\exb$ shear is increased, the amplitude of the saturated potential is reduced at the largest scales (lowest wave numbers). In \QLK{}, this is due to a shift of the maximum of the saturated potential towards smaller scales corresponding to the usual picture of the non-linear effect of the $\exb$ shear. In \GKW{}, a flattening of the saturated potential amplitude is rather observed around its maximum. Both codes exhibit a weak dependence of their saturated potential with $\gamma_E$ at $\kthe\rho_s>0.2$. Quantitatively, in \QLK{}, the reduction of the saturated potential maximum amplitude is underestimated at lower $\exb$ and overestimated at higher $\exb$ shear values. Despite these quantitative differences, the non-linear fluxes quenching with $\exb$ shear is captured qualitatively with a shifted eigenfunction calculated in the fluid limit. In the next section, the quasi linear fluxes are compared to non-linear simulations and the influence of the saturated potential of the fluxes is further discussed.

\subsection{Comparison of \QLK{} fluxes with non-linear simulations}
\label{subsec:comp flux}
\begin{figure*}%
  \subfigure{\includegraphics[width=0.5\textwidth]{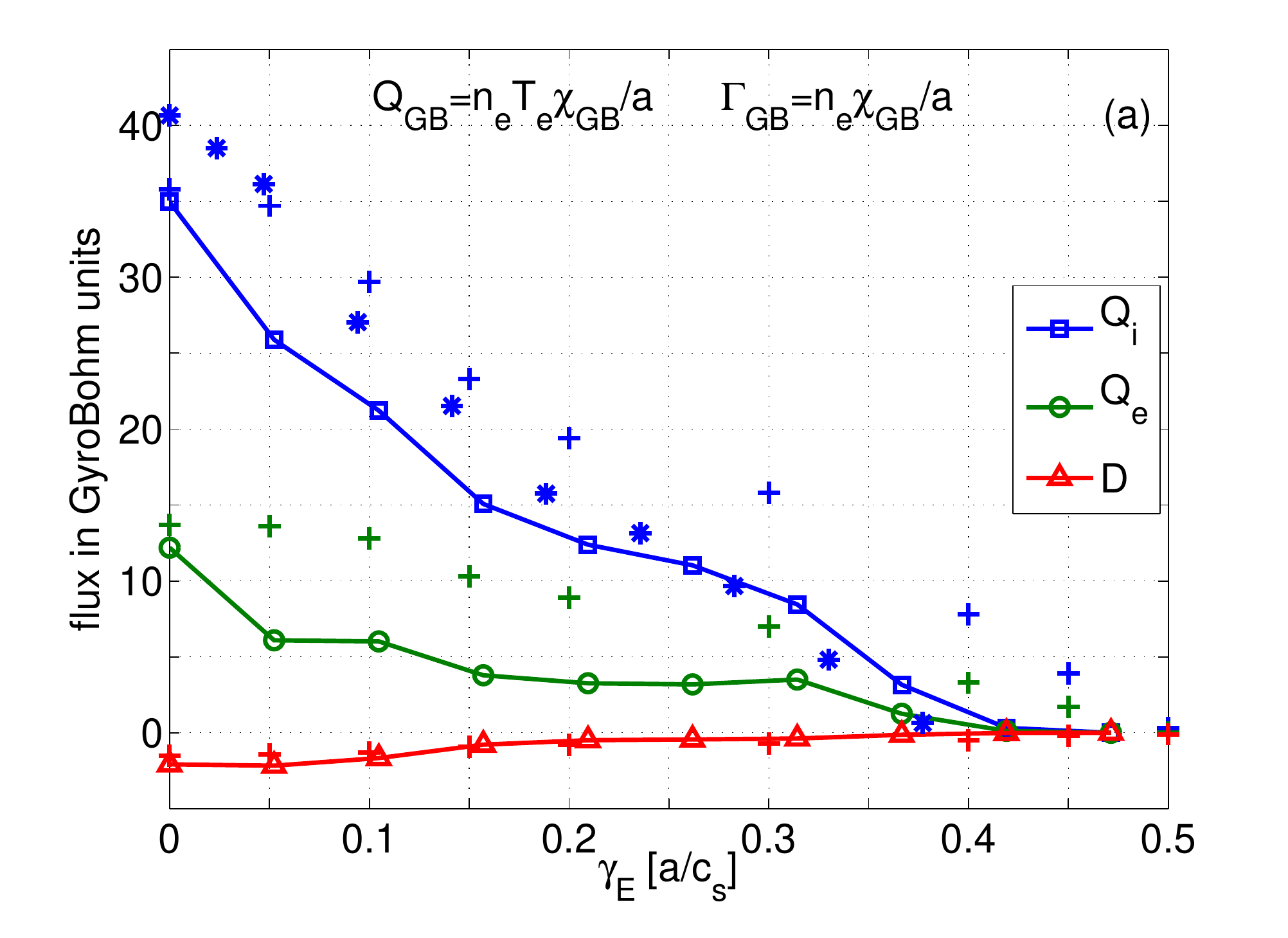}
  \label{fig:comp_gyro_Qie_D_scangamE}}%
	\subfigure{\includegraphics[width=0.5\textwidth]{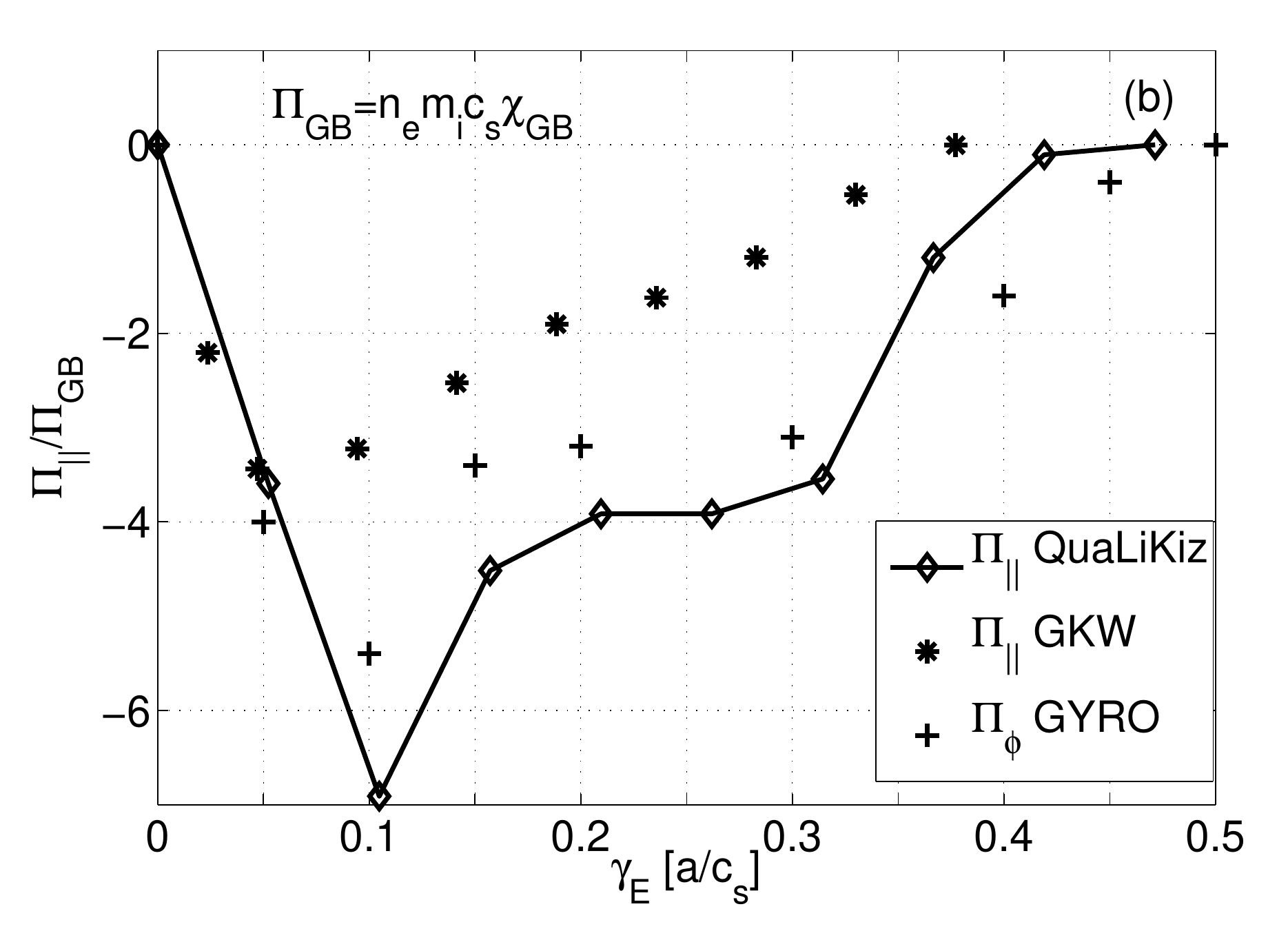}
	\label{fig:comp_gyro_Pi_scangamE}}%
		\caption{(a) Ion and electron heat fluxes, particle flux and (b) angular momentum flux for GA-std parameters. Here $\Pi_\parallel\equiv \Pi_{RS}$ since $\upar=\gradu=0$. The solid lines are \QLK{} results, the stars $\mathbf{*}$ are \GKW{} data from \protect{\citep{cass09}} and the crosses $\mathbf{+}$ are \gyro{} data from \protect{\citep{staeb13}}. $a/c_s$ units have to be multiplied by $3/\sqrt{2}$ to have their $R/v_{Ti}$ equivalent.}
\label{fig:comp_gyro_fluxes_scangamE}
\end{figure*}

To finally evaluate the model presented above, the resulting heat, particle and momentum fluxes are compared to non linear simulations. First, the impact of $\exb$ shear alone is studied in Figure~\ref{fig:comp_gyro_fluxes_scangamE}, i.e. $\upar$ and $\gradu$ are artificially set to 0. GA std case parameters are used to compare \QLK{} predictions with published results from non-linear \gyro{} \citep{staeb13} and \GKW\citep{cass09}.  

\QLK{} heat and particle fluxes are smoothly reduced and quenched for $\gamma_E>0.4c_s/a$ as illustrated in Figure~\ref{fig:comp_gyro_Qie_D_scangamE}. This quench value is lower than what is found by \gyro{} simulations \citep[see][Figure~1]{staeb13} but is in agreement with the value obtained with \GKW{} \citep[see][Table II]{cass09} using non-linear \GKW{}. \QLK{} predictions for the fluxes amplitude lies between non-linear \GKW{} and non-linear \gyro{} for the ion heat flux. In \gyro{} the fluxes reduction with increasing $\exb$ shear is notably slower than found with \GKW{} and \QLK{} as illustrated by Figure~\ref{fig:comp_gyro_Qie_D_scangamE}.

The angular momentum flux $\Pi_\parallel$ is presented in Figure~\ref{fig:comp_gyro_Pi_scangamE}. As $\upar$ and $\gradu$ are set to zero, $\Pi_\parallel$ corresponds to the residual stress $\Pi_{RS}$ in this case. In absolute value, the momentum flux increases at first with $\gamma_E$ due the $\exb$ shear asymmetrization of the eigenfunction. Then, the momentum flux is slowly reduced due to the turbulence quenching by the $\exb$ shear. This qualitative trend is in agreement with non-linear simulations. 
Quantitatively, \QLK{} overestimates the momentum flux found with \GKW{} by $\sim 50\%$ but is in agreement with \gyro{} simulations. \gyro{} was run with a circular Miller equilibrium retaining the finite $\epsilon$ effects, which are not present in the \GKW{} simulations with the $\hat{s}-\alpha$ equilibrium nor in \QLK. The discrepancy between \QLK{} and \GKW{} is related to the overestimation of the saturated potential amplitude at lower $\kthe\rho_s$ and intermediate values of $\gamma_E$ in \QLK{} detailed in the previous section. This is a necessary trade-off to estimate the $\exb$ shear induced turbulence quench and momentum flux in a reduced model compatible with the integrated modeling framework without using any fitting model. It is interesting to note that a fluid model captures the essential physical mechanisms of the complex $\exb$ shear action on the modes.

Now, the effect of $\gradu$ and $\upar$ on the momentum flux are analyzed. To perform this analysis, the following non dimensional quantities are employed: The Prandtl number $\frac{\chi_\parallel}{\chi_i}$ and the pinch number $\frac{RV_\parallel}{\chi_\parallel}$. They facilitate the comparison with non linear simulations as the saturated potential does not appear in these ratio. 

In \QLK{}, isolating conductive and convective contributions to the momentum flux is not straightforward due to $\x_0$ dependencies presented in Sec.~\ref{subsec:mom_flux_exp}. To evaluate the different parts of the momentum flux, the 3-point presented in Sec.~\ref{subsec:mom_flux_exp} can be simplified when dealing with test cases. A simulation with only $\gradu$ as a symmetry breaker ($\upar=\gamma_E=0$) is performed. The ratio of the momentum flux to the ion heat flux then gives the Prandtl number. To evaluate the \emph{total} convective part, a simulation with only $\upar$ -- $\gradu=\gamma_E=0$ -- is carried out. The ratio between the resulting momentum flux to the previous $\gradu$-only momentum flux gives the pinch number. In the following, this method is called \emph{2-point method}. Compared to the 3-point method, the modification of the conductivity by $\exb$ shearing (through the force balance equation) is neglected. Indeed, $\gamma_E$ is artificially put to 0 as is usually done in momentum diffusivity/pinch analysis with non-linear gyro-kinetic simulations \citep{peet11}. 

\begin{figure*}%
 \subfigure{\includegraphics[width=0.5\textwidth]{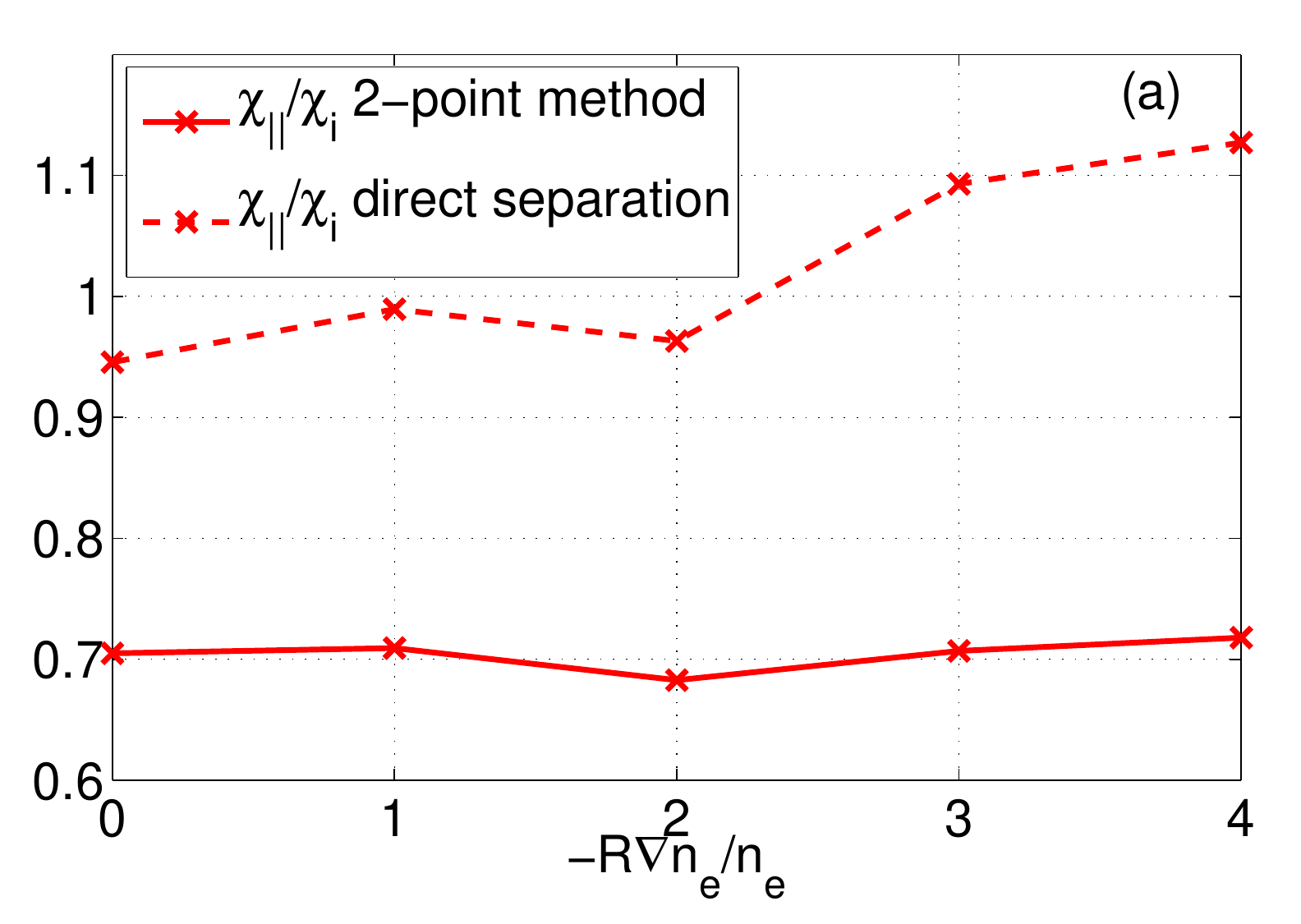}
 % \caption{$\upar$, $\gradu$ and $\mathrm{x}_0$ contributions to the momentum flux. }
  \label{fig:Pr_scan_An}}
	\subfigure{\includegraphics[width=0.5\textwidth]{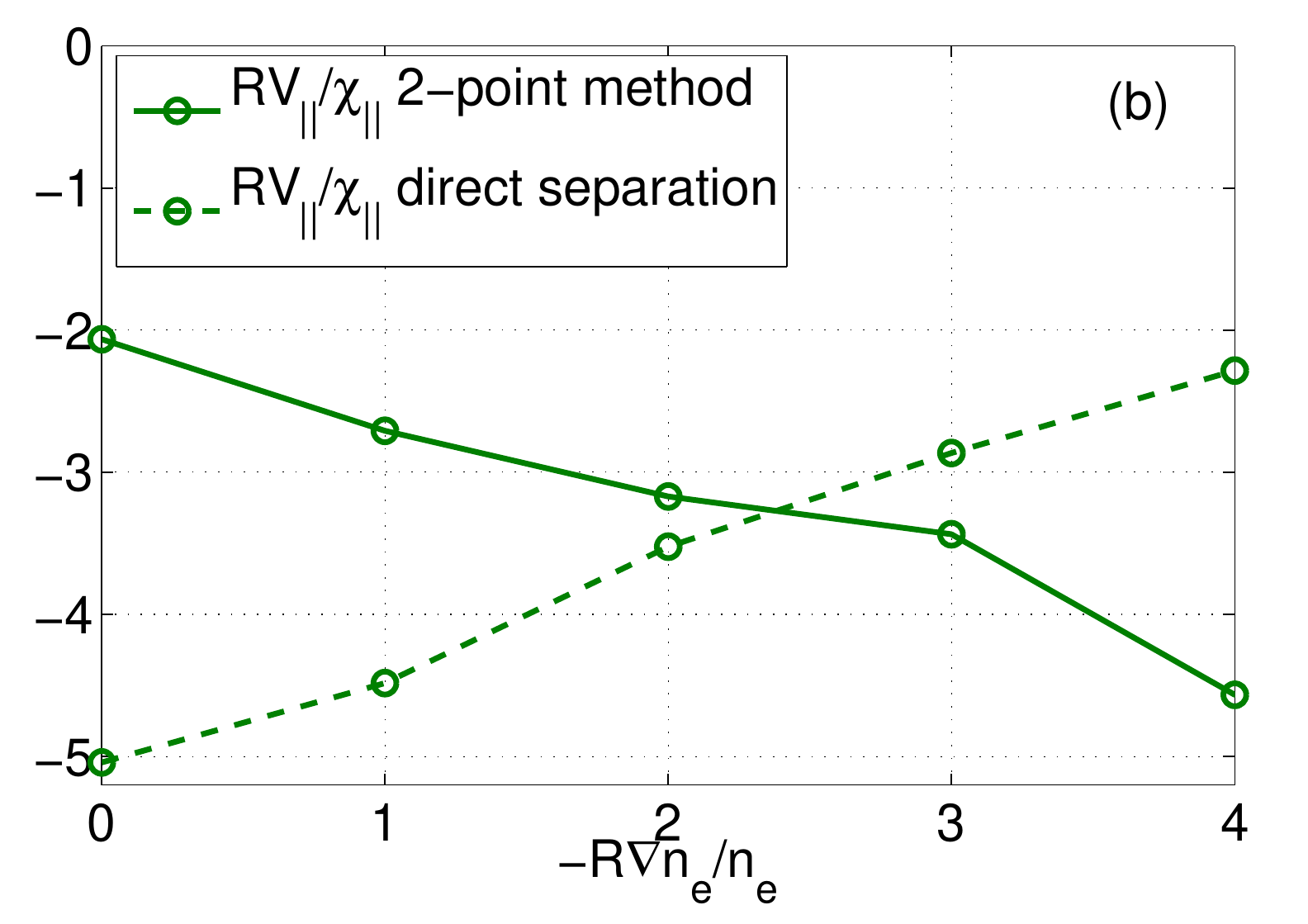}
%   \caption{$\Pi_{\x0}$ compared with the momentum flux (plain) and the residual stress (diamonds).}
  \label{fig:pinch_scan_An}}
\caption{(a) Prandtl (red crosses) and (b) pinch number (green circles) calculated with the \emph{direct separation method} (dashed curves) and with the \emph{2-point method}\protect{\citep{peet11}} (plain curves)}
\label{fig:Pr_pinch_scan_An}
\end{figure*}

Two \QLK{} simulations based on GA-std case parameter set are performed for the validation of the conductive and convective contributions to the momentum flux calculated by the 2-point method:
\begin{itemize}
\item one with $\frac{-R\gradu}{v_{Ti}}=1$, $\frac{\upar}{v_{Ti}}=0$;
\item one with $\frac{-R\gradu}{v_{Ti}}=0$, $\frac{\upar}{v_{Ti}}=0.2$.
\end{itemize}
As explained in Sec.~\ref{subsec:mom_flux_exp}, a direct extraction of a $\Pi_{\nabla{u}}$ and a $\Pi_u$ -- corresponding to diffusive and convective contributions to the momentum transport \emph{without} taking the eigenfunction shift effect into account -- is possible in \QLK{}. This method called \emph{direct separation method} is compared to the 2-point method in Figure~\ref{fig:Pr_pinch_scan_An} to give an idea of the impact of the eigenfunction shift on $\chi_\parallel$ and $V_\parallel$.

The normalized density gradient $R/L_n$ was varied from 0 to 4. Indeed, results from non-linear gyrokinetic simulations indicate a strong correlation between $R/L_n$ and the pinch number \citep{hahm07,peet11}, the Prandtl number being weakly correlated. In Figure~\ref{fig:Pr_pinch_scan_An}, the Prandtl number is displayed with crosses and the pinch number with circles, the results from the 2-point method being in plain curves and the estimations via direct separation in dashed curves. 

The Prandtl number deduced from the 2-point method is found to be close to 0.7 agreeing with quasi-linear \citep{peet05} and non-linear simulations\citep{peet11}. Due to the omission of the eigenfunction shift effect, the direct separation in \QLK{} gives a higher Prandtl number, close to one, as predicted in early theoretical calculations\citep{diam88}.
%\begin{figure}
%\label{fig:Pr_pinch_scan_An}
%\centering
%\includegraphics[width=0.5\textwidth]{Pr_pinch_scanAn.pdf}%\hfill
%\caption{Prandtl (red crosses) and pinch number (green circles) calculated with the \emph{direct separation method} (dashed curves) and with the \emph{2-point method}\protect{\citep{peet11}} (plain curves)}
%\end{figure}
Using the 2-point method, the pinch number $\frac{RV_\parallel}{\chi_\parallel}$ is found to vary from $-2$ to $-5$, with a strong correlation with $R/L_n$, as in \citep{peet11}. When neglecting the eigenfunction shift effects, i.e. with the direct separation technique, the correlation with $R/L_n$ is inverted. Taking the ratio of the momentum fluxes amplifies the error. This illustrates that the eigenfunction shift has to be taken into account to have the correct dependencies and values of the momentum flux. 

To summarize this section, the quasi-linear momentum flux derived in Sec.~\ref{subsec:mom_flux_exp} was successfully benchmarked against non-linear simulations, including the momentum diffusivity, the momentum pinch and the residual stress. For the conductive and convective parts of the momentum flux, two methods were presented and compared. The importance of the eigenfunction shift contribution was illustrated. In the next section, the influence of the $\exb$ shear on the momentum flux will be analyzed with \QLK{} and compared to the experimental results.

\section{Comparison with the experiment}
\label{sec:exp}
In this final section, a JET H-mode shot is analyzed with \QLK{}. The Prandtl and pinch numbers are found compatible with the experiment on a large part of the radius. However the effective ion heat flux is significantly compared to the experimental value from JETTO in interpretative mode. 

\begin{figure}
\includegraphics[width=\textwidth]{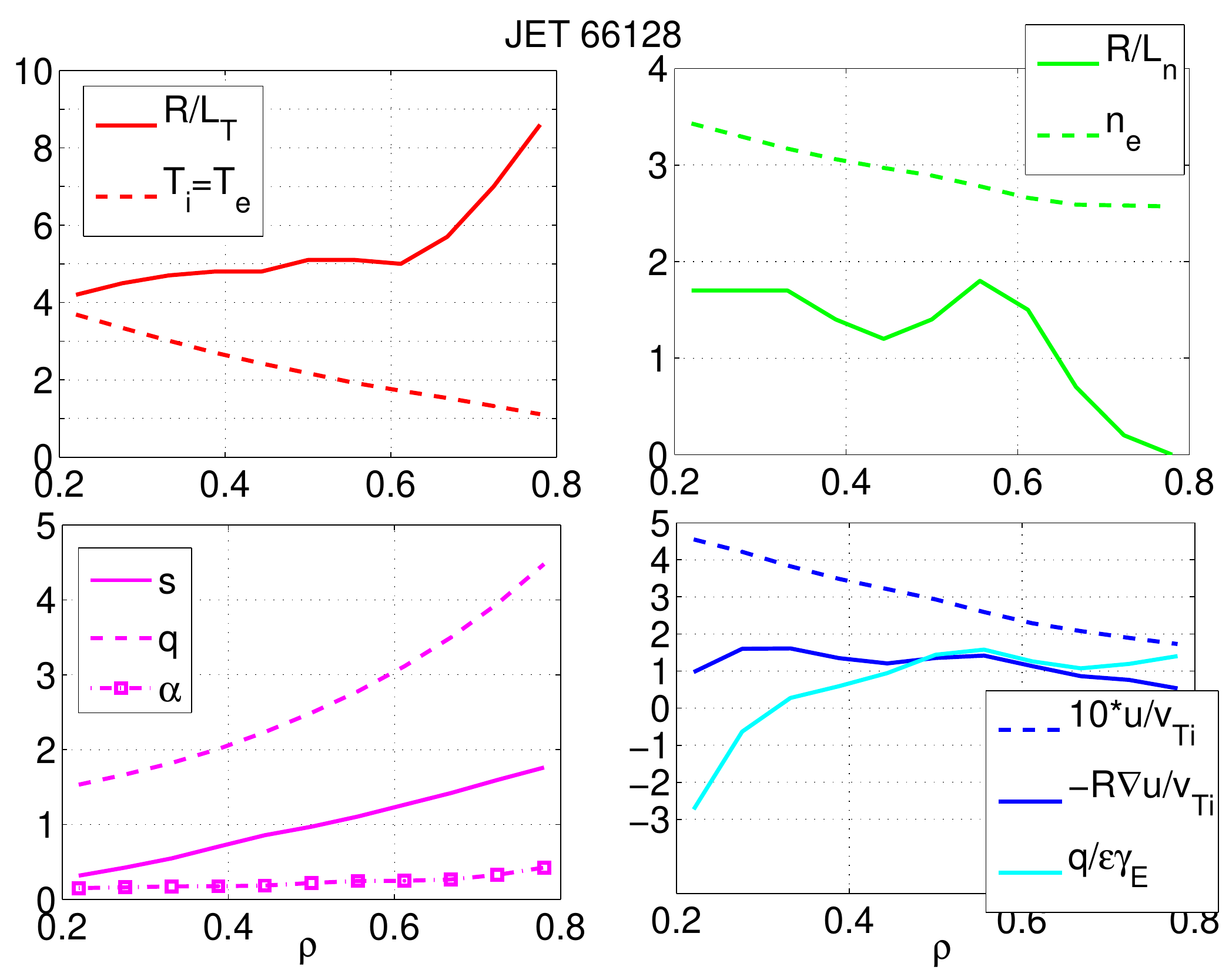}
\caption{\label{fig:66128_parameters}Input parameters for \QLK{} simulation of JET shot 66128. All parameters were taken from JETTO fit realized for \GKW{} simulations of \citep{tala09} except $T_i=T_e$. $Z_\textsl{eff}=2$}
\end{figure}

The analyzed shot, from Tala et al.\citep{tala09}, is an NBI modulation experiment proving the experimental evidence of a momentum pinch. To evidence the presence of a momentum pinch, the amplitude and phase of the modulated toroidal velocity was simulated with JETTO:
\begin{itemize}
\item either with only momentum diffusivity \textsl{i.e.} $\chi_\phi/\chi_i=\chi_{\phi,\textsl{eff}}/\chi_i\approx0.25$ 
\item or with both momentum diffusivity and pinch. $\chi_\phi/\chi_i=1$ matching theory based estimations \citep{diam88} in older calculations\citep{tala07}, or computed with gyrokinetic simulations \citep{tala09}, and $v_{pinch}\approx15$m/s adapted to match the experimental effective diffusivity $\chi_{\phi,\textsl{eff}}$ or, equivalently, the modulated toroidal velocity amplitude.
\end{itemize}
Tala et al.\citep{tala07,tala09} showed that both the amplitude and the phase of the experimental toroidal velocity  are only correctly reproduced when a momentum pinch is taken into account. However, the residual stress was neglected in their analysis. Quasi-linear gyrokinetic simulations are performed with \QLK{}. The global parameters are the ones used in \GKW{} for Figure~3 of \citep{tala09}. The main input parameters of the simulation are displayed in Figure~\ref{fig:66128_parameters}. All parameters are taken from JETTO interpretative run performed for \GKW{} simulation of \citep{tala09} with the exception of $T_i=T_e$ as there is no evidence from the CX and ECE signals for $T_i\neq T_e$. Since \QLK{} has a circular equilibrium the gradients are averaged over the flux surface.

\begin{figure*}
\centering
\subfigure{\includegraphics[width=0.48\textwidth]{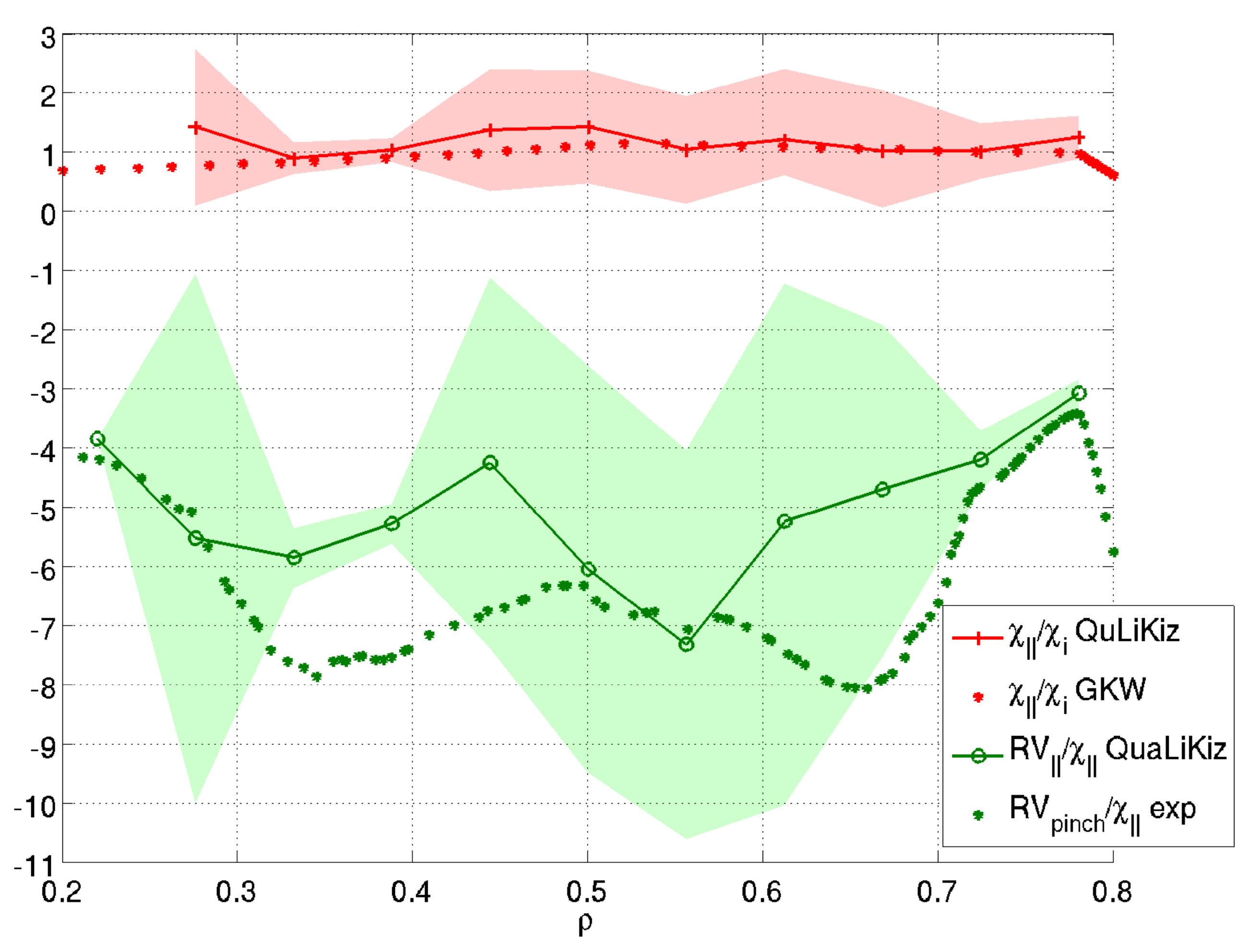}\label{fig:Pr_pinch_66128}}
\subfigure{\includegraphics[width=0.48\textwidth]{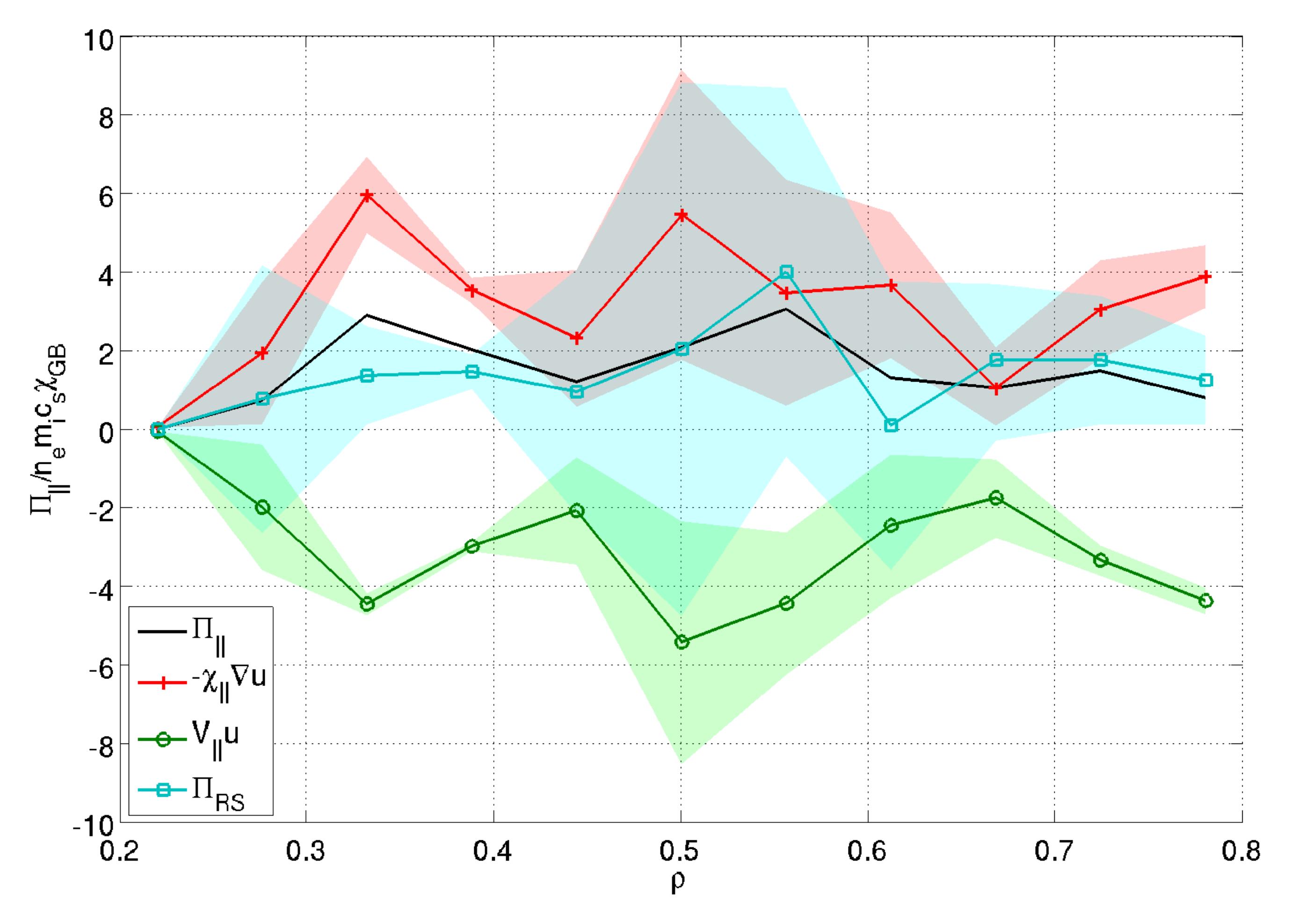}\label{fig:mom_fluxdetail66128}}
\caption{Left: Prandtl number (red crossed) and pinch number(green circles) calculated by a 3-point method. Right: Detail of the different contributions to $\Pi_{||}$.}
 \end{figure*}

The $\exb$ shear calculated with the radial force balance equation on the carbon impurity is significant in this shot, as indicated in Figure~\ref{fig:66128_parameters}. Since the collisionality is weak in this shot --- $\nu^*\in [0.03; 0.08]$ --- the neoclassical value for the poloidal velocity is given by the banana regime value $v_{\theta,C}B_\varphi=1.17\nabla{T_C}/{6e}$. The \emph{3-point method} presented in Sec.~\ref{subsec:mom_flux_exp} is used to correctly account for the different contributions to the momentum flux and quantify the momentum diffusivity, the momentum pinch and the residual stress. As indicated in Sec.~\ref{subsec:mom_flux_exp}, a simulation is performed with the experimental conditions described in Figure~\ref{fig:66128_parameters}, one is performed with the parallel velocity modified by $\pm 20\%$ with the corresponding modification in $\gamma_E$ and $\gradu$ and one with the parallel velocity \emph{incremented} by $\pm0.05V_{Ti}/R$ with the corresponding modification of $\gamma_E$ but no change in $\gradu$. The resulting Prandtl and pinch numbers are given in Figure~\ref{fig:Pr_pinch_66128}. The colored regions in this plot corresponds to the uncertainties linked to the linearization performed to extract these numbers. They are calculated by performing 5 simulations with different modifications of the velocity and combining the results.

The estimated Prandtl number lies within 0.8 and 1.4, close to \GKW{} predictions used in \cite{tala09}. The pinch number calculated with \QLK{} ranges from 3 to 7, in good agreement with the experimental values ranging from 3 to 8. The large uncertainties obtained with the 3-point method indicates that the momentum flux changes in a complex way with $\upar$ and $\gradu$ which the linearization employed to get Figure~\ref{fig:Pr_pinch_66128} does not reflect.

The contributions to the momentum flux from $\upar$, $\gradu$ and the residual stress are compared in Figure~\ref{fig:mom_fluxdetail66128}. The estimated residual stress seems not entirely negligible in this shot. However a definitive conclusion would require smaller error bars. Moreover some significant contributions to the residual stress are not taken into account in local models such as \QLK{} as pointed out by \citep{waltz11}.

\begin{figure}
\centering
\includegraphics[width=0.5\textwidth]{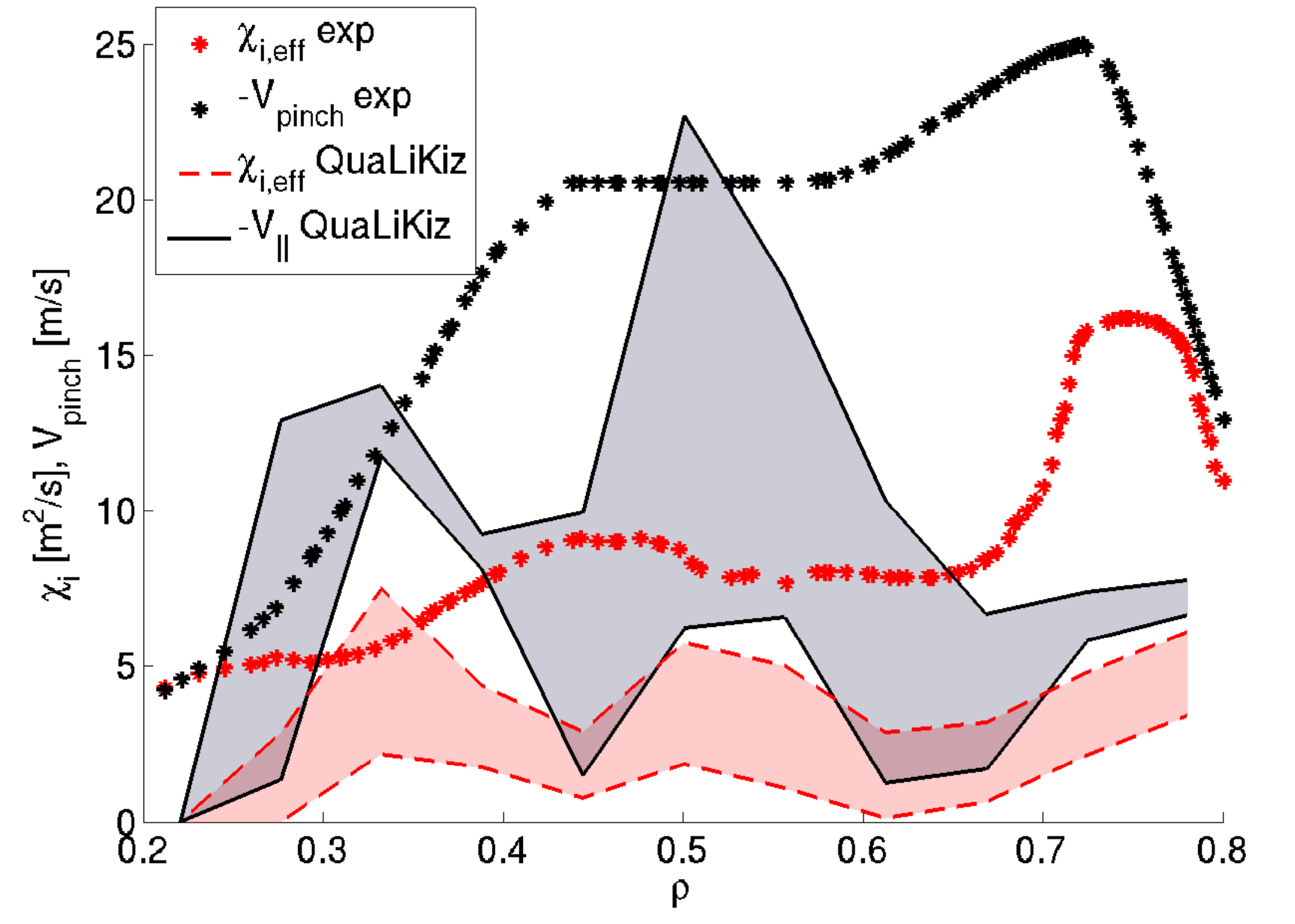}
\caption{Ion heat flux diffusivity (red dashed) and pinch velocity (plain). The colored regions correspond to a 20\% variation of $R/L_T$ with associated variation of $\gamma_E$.}
\label{fig:66128_comp_vpinch_chi}
\end{figure}

Finally ,the pinch velocity itself $-V_\parallel$ (plain curve) is plotted along with the effective ion heat flux $\chi_{i,\textsl{eff}}$ (dashed curve) in Figure~\ref{fig:66128_comp_vpinch_chi} and compared to the experimental estimates. To improve the robustness of the results and reproduce experimental uncertainties, $R/L_T$ was varied by 20\% with the associated modification of $\gamma_E$. It corresponds to the colored regions of Figure~\ref{fig:66128_comp_vpinch_chi}. Even when increasing the temperature gradients by $20\%$ $\chi_{i,\textsl{eff}}$ is underestimated compared to the experiment. This advocates for including a more refined magnetic equilibrium in \QLK. Indeed, averaging over the flux surface is a way to take the stabilizing effect of the elongation into account. However, it appears that the stabilization is overestimated by this method. Increasing by 20\% the gradients gives a closer results. This is equivalent to taken the gradient at the midplane. According to the good agreement on the pinch and the Prandtl number, $V_\parallel$ is also underestimated in \QLK. 
Outside $\rho=0.5$, the discrepancy between \QLK{} and JETTO predictions enlarges. This may comes from the choice of $T_e=T_i$ made in \QLK{} simulations based on CX and ECE signals in disagreement with JETTO fit. However, the fact that JETTO runs fail to reproduce the experimental phase of the modulated velocity at this radii is worth noticing.

To summarize, considering the experimental uncertainties on the various gradients used as inputs, \QLK{} estimations of the Prandtl number and the momentum pinch are close enough to the ones evaluated from the experiment. In particular, an inward convective flux of momentum is found in the model and the experiment with a pinch number ranging from 5 to 8. However, a quantitative analysis on the fluxes intensity remains difficult due to the fluxes sensitivity to the temperature gradients combined with the difficulty to measure them accurately in experiments. To overcome this issue, the present version of \QLK{} has to be coupled to an integrated platform such as CRONOS. This will enable driving \QLK{} via the sources which is more relevant physically than to impose the gradients.

\section{Conclusions}
With the aim to improve and broaden the capabilities of first principle based transport models for integrated modeling, the gyrokinetic transport code \QLK{}\citep{bour07,casati09,citrin12} has been upgraded to include sheared flow effects and momentum flux calculation. 

For momentum studies, the shape of the eigenfunctions in the parallel direction is essential as illustrated in section~\ref{subsec:comp flux}. The reduced fluid model used for \QLK{} eigenfunctions was shown to recover the correct dependencies with the parallel velocity, its gradient and the $\exb$ shear without any ``free-fitting parameters'' even close to the turbulent threshold; by direct comparisons with self-consistent gyrokinetic eigenfunctions for $\upar$ and $\gradu$. Recovering the low-$k$ turbulence quench, the heat and particle fluxes reduction and the residual stress induced by $\exb$ shear from non-linear simulations demonstrated that $\exb$ shear modeling is valid as well. The results on the residual stress remarkably showed that a shift of the linear eigenfunctions is enough to get the correct effect of the $\exb$ shear on the saturated potential with a mixing length rule. 

Separating the different contributions from $\upar$, $\gradu$ and $\exb$ shear to the momentum flux appeared to be challenging. With a \emph{3-point method}, the momentum diffusivity and pinch and the $\exb$ induced residual stress can be calculated. The Prandtl and pinch numbers calculated this way showed good agreement with both non-linear simulations and NBI modulation experimental results. In particular, the correlation of the pinch number with $R/L_n$ was recovered. The residual stress was evaluated but no definitive conclusions should be drawn due to the uncertainties linked to the 3-point method i.e. the total flux is not linear in $\upar$, $\gradu$, $\gamma_E$. The insight gained by analyzing experiments dedicated to the residual stress characterization appears limited in the local approach taken in \QLK{}. Since the residual stress is a higher $\rho^*$ quantity, it cannot be properly determined by local simulations. 

From NBI modulation experiments, the variability of \QLK{} predictions within experimental uncertainties was underlined, pointing out the limitations of gradient driven simulations for comparisons with experiments.

Finally, with the new features presented in this paper, \QLK{} opens the way for simulating consistently $T_e$, $T_i$, $n_e$ and $v_\parallel$ profiles in integrated modeling platforms such as CRONOS. This will have the side benefit of driving \QLK{} with the sources instead of imposing the gradients, improving its prediction capabilities.

\section*{Acknowledgments}
The authors wish to acknowledge fruitful discussions and useful comments from J. Citrin, G. Dif-Pradalier, N. Fedorczak and Y. Sarazin.
This work was granted access to the National Research Scientific Computing Center resources, supported by the Office of Science of the U.S. Department of Energy under Contract No. DE-AC02-05CH11231. The authors are very grateful to D. Mikkelsen for having provided computational resources.
This work, supported by the European Communities under the contract of Association between EURATOM and CEA, was carried out within the framework of the European Fusion Development Agreement. The views and opinions expressed herein do not necessarily reflect those of the European Commission. 

\appendix
\section{Passing particle functional}
\label{subsec:passing}
Before integration the passing particle functional reads:
\begin{align}\label{eq:I_pass}\begin{split}
\mathcal{I}_{s,pass}=&\sum_{\epsilon_\parallel=\pm 1}\left(1+\frac{2\upar}{v_{Ts}}\epsilon_{\|}\sqrt{\xi(1-\lambda b)}+\frac{u_{\|}^2}{v_{Ts}^2}\left(2\xi(1-\lambda b)-1\right)\right)\\
&\frac{\frac{R}{L_{Ts}}^*\xi+2(\frac{R}{L_u}-\frac{R}{L_{Ts}})\frac{u_\|}{v_{Ts}}\epsilon_\|\sqrt{\xi(1-\lambda b)}+\frac{R}{L_{ns}}-\frac{3}{2}\frac{R}{L_{Ts}}+\frac{u_\|^2}{v_{Ts}^2}(\frac{R}{L_{Ts}}-2\frac{R}{L_u})-\frac{\varpi}{n\bar\omega_{ds}}}{(2-\lambda b)f_\theta\xi+\epsilon_{\|}\frac{x}{d}\frac{\omega_b}{n\bar\omega_{ds}}-\frac{\varpi}{n\bar\omega_{ds}}+\imath o^+}
\end{split}\end{align}
The integration over $\lambda$ and $\xi$ is then performed. In \QLK, the integration over $\lambda$, not tractable analytically, is simplified. It is considered that the passing particle pitch-angle variation does not influence the drift frequencies so that they can be averaged over $\lambda$. This assumption is correct for the curvature and $\nabla{B}$ drift for which the pitch angle variation represents no more than 50\% of its value. For $\kpar\vpar$ expression however, this means that its value will be overestimated for barely passing particles. The result is given in (~\ref{eq:passing}) using the Fried-Conte function $\ds{Z(z)=\frac{1}{\sqrt{\pi}}\int_ {-\infty}^{+\infty}\frac{e^{-v^2}}{v-z}\mathrm{d}v}$.
\begin{align}\label{eq:passing}
\begin{split}
\left\langle\mathcal{I}_{s,pass}\right\rangle_p=\frac{3f_p}{2f_\theta}&\bigg[\frac{R}{L_{Ts}}\frac{Z_2(V_+)-Z_2(V_-)}{V_+-V_-}+\left(\frac{R}{L_{ns}}-\frac{3}{2}\frac{R}{L_{Ts}}-\frac{\varpi}{n\bar\omega_{ds}}\right)\frac{Z_1(V_+)-Z_1(V_-)}{V_+-V_-}\bigg]\\
+\frac{3f_p}{f_\theta}&\bigg[\frac{\upar}{v_{Ts}}\frac{R}{L_{Ts}}\frac{V_+Z_2(V_+)-V_-Z_2(V_-)}{V_+-V_-}+
\left(\frac{R}{L_u}+\frac{\upar}{v_{Ts}}\left(\frac{R}{L_{ns}}-\frac{5}{2}\frac{R}{L_{Ts}}-\frac{\varpi}{n\bar\omega_{ds}}\right)\right)\frac{V_+Z_1(V_+)-V_-Z_1(V_-)}{V_+-V_-}\bigg]\\
+\frac{f_p}{f_\theta}\frac{\upar}{v_{Ts}}&\bigg[A_T\frac{\upar}{v_{Ts}}\frac{Z_3(V_+)-Z_3(V_-)}{V_+-V_-}+\left(2\frac{R}{L_u}
\frac{\upar}{v_{Ts}}\left(\frac{R}{L_{ns}}-\frac{7}{2}\frac{R}{L_{Ts}}-\frac{\varpi}{n\bar\omega_{ds}}\right)\right)\frac{Z_2(V_+)-Z_2(V_-)}{V_+-V_-}\bigg]\\
-\frac{3f_p}{f_\theta}\frac{\upar}{v_{Ts}}&\left[\frac{R}{L_{Ts}}\frac{\upar}{v_{Ts}}\frac{Z_2(V_+)-Z_2(V_-)}{V_+-V_-}+\left(2\frac{R}{L_u}\frac{\upar}{v_{Ts}}\left(\frac{R}{L_{ns}}-\frac{5}{2}\frac{R}{L_{Ts}}-\frac{\varpi}{n\bar\omega_{ds}}\right)\right)\frac{Z_1(V_+)-Z_1(V_-)}{V_+-V_-}\right]
\end{split}\end{align}
where $f_p$ is the passing particle fraction. $Z_1$, $Z_2$ and $Z_3$ are defined are based on the Fried-Conte function $Z$: $Z_1(z)=z+z^2Z(z)$, $Z_2(z)=\frac{1}{2}z+z^2Z_1(z)$ and $Z_3(z)=\frac{3}{4}z+z^2Z_2(z)$.
The variables $V_+$ and $V_-$ correspond to the poles of (\ref{eq:I_pass}). They are defined by:
\begin{align}\label{eq:poles}
\begin{split}
V_\pm&=\frac{1}{2}\frac{v_{Ts}x}{qRd}\frac{\bar\omega_b}{f_\theta n\bar\omega_{ds}}\pm\sqrt{\Delta}\\
\Delta&=\left(\frac{1}{2}\frac{v_{T_s}x}{qRd}\frac{\bar\omega_b}{f_\theta n\bar\omega_{ds}}\right)^2 +\frac{\varpi}{f_\theta n\bar\omega_{ds}}
\end{split}\end{align}
The integration over $k_r$ remains to be performed. As expressed in (\ref{eq:kparvpar}), there remain some $x$ dependence in the passing particle functional. Moreover, $\varpi=\omega-n\omega_{E \times B}$ contains an $x$ dependence too. To take all effects into account, an integration over $k_r$ and $x=r-r_0$, where $x\ll r_0$, is performed as derived by Garbet et al.\citep{garb90} and presented first in App. A.4.2 of \citep{bour02} for \QLK{} framework. The expression of $\mathcal{L}_{s,pas.}=\int_{-\infty}^{+\infty}\frac{\mathrm{d}k_r}{2\pi}\left\langle\I_{s,pass}\right\rangle_p\mathcal{B}_0(k_\theta\rho_s)$ is then transformed into:
\begin{equation}
\mathcal{L}_{s,pas.}=\int_{-\infty}^\infty\frac{\mathrm{d}k_+}{2\pi}\iint_{-\infty}^\infty\mathrm{d}x_+\mathrm{d}x_-\tilde\phi(x_+-\frac{x_-}{2})\tilde\phi^*(x_++\frac{x_-}{2})e^{\imath k_+x_-}\left\langle\I_{s,pass}\right\rangle_{p}\mathcal{B}_0(k_\theta\rho_s)
\end{equation}
As shown in section \ref{sec:fluid} -- in presence of $\upar$, $\gradu$ and $\exb$ shear -- $\tilde\phi(x)$ is a shifted Gaussian:
\begin{equation}
\tilde\phi(x)=\phi_0\exp(-\frac{(x-\x_0)^2}{2w^2})
\end{equation}
Therefore, the product $\phi\phi^*$ can be written as:
\begin{equation}
\tilde\phi\tilde\phi^*=\phi_0^2\exp\left(-\frac{(x_+-\Re(\x_0)-k_+\Im(\w^2))^2}{\Re(\w^2)}-\Re(\w^2)\left(k_+-\frac{\Im(\x_0)}{\Re(\w^2)}\right)^2\right)
\end{equation}
Dimensionless quantities $\rho^*$ and $k^*$ are defined for the integration over $x_+$ and $k_+$:
\begin{align}
\begin{split}
\rho^{*2}&=\frac{(x_+-\Re(\x_0)-k_+\Im(\w^2))^2}{\Re(\w^2)}\\
k^{*2}&=\Re(\w^2)\left(k_+-\frac{\Im(\x_0)}{\Re(\w^2)}\right)^2
\end{split}
\end{align}
In (\ref{eq:poles}), $x$ is replaced by $\ds{\rho^*\sqrt{\Re(\w^2)}+\Re(\x_0)+k\Im(\w^2)}$ and $\ds{k=\frac{k^*}{\sqrt{\Re(\w^2)}}+\frac{\Im(\w^2)}{\Re(\w^2)}}$, $\Re(\w^2)$ being defined positive which ensures $\left|\tilde\phi\right|^2$ is finite.
The passing particle functional then become:
\begin{equation}
\mathcal{L}_{s,pass}=\int_{-\infty}^\infty\frac{\mathrm{d}k^*}{\sqrt{\pi}}e^{-k^{*2}}\int_{-\infty}^\infty\frac{\mathrm{d}\rho^*}{\sqrt{\pi}}e^{-\rho^{*2}}\left\langle\mathcal{I}_{s,pass}\right\rangle_p(k^*,\rho^*)\mathcal{B}_0(k_\theta\rho_s)
\end{equation}

\section{Trapped particle functionals}
\label{subsec:trapped}
For trapped particles, there are no $\theta$ dependence in the drifts, since the bounce average is performed. $\kpar\vpar$ is therefore expressed in terms of the poloidal wave number $m$:
\begin{equation}
\kpar\vpar=\pm\frac{mv_{Ts}}{qR}\sqrt{\xi(1-\lambda b)}
\end{equation}
It is also stressed that no assumption is taken on $\lambda$.  
\begin{align}\label{eq:I_tr}\begin{split}
\mathcal{I}_{i,m,tr}=&\sum_{\epsilon_\parallel=\pm 1}\left(1+\frac{2\upar}{v_{Ts}}\epsilon_{\|}\sqrt{\xi(1-\lambda b)}+\frac{u_{\|}^2}{v_{Ts}^2}\left(2\xi(1-\lambda b)-1\right)\right)\\
&\frac{\frac{R}{L_{Ts}}^*\xi+2(\frac{R}{L_u}-\frac{R}{L_{Ts}})\frac{u_\|}{v_{Ts}}\epsilon_\|\sqrt{\xi(1-\lambda b)}+\frac{R}{L_{ns}}-\frac{3}{2}\frac{R}{L_{Ts}}+\frac{u_\|^2}{v_{Ts}^2}(\frac{R}{L_{Ts}}-2\frac{R}{L_u})-\frac{\varpi}{n\bar\omega_{ds}}}{(2-\lambda b)f_\theta\xi+\epsilon_{\|}m\frac{\omega_b}{n\bar\omega_{ds}}-\frac{\varpi}{n\bar\omega_{ds}}+\imath o^+}
\end{split}\end{align}
The attentive reader noticed that (\ref{eq:I_tr}) is expressed for trapped ions. Its expression is different for trapped electrons because electron-ion collisions are integrated in \QLK. Since the effect of collisionality is most important on trapped electrons\citep{Connor06}, collisions are only implemented in trapped electron functionals as detailed in \citep{roma07}. For $\mathcal{I}_{e,m,tr}$, $\nu_{ie}$ is included in (\ref{eq:I_tr}) in place of the Landau prescription for causality, the small quantity $\imath o^+$, through a Krook operator presented in \citep{roma07}. The expression of $\mathcal{I}_{e,m,tr}$ is 
\begin{align}\label{eq:I_e,tr}\begin{split}
\mathcal{I}_{e,m,tr}=&\sum_{\epsilon_\parallel=\pm 1}\left(1+\frac{2\upar}{v_{Ts}}\epsilon_{\|}\sqrt{\xi(1-\lambda b)}+\frac{u_{\|}^2}{v_{Ts}^2}\left(2\xi(1-\lambda b)-1\right)\right)\\
&\frac{\frac{R}{L_{Ts}}^*\xi+2(\frac{R}{L_u}-\frac{R}{L_{Ts}})\frac{u_\|}{v_{Ts}}\epsilon_\|\sqrt{\xi(1-\lambda b)}+\frac{R}{L_{ns}}-\frac{3}{2}\frac{R}{L_{Ts}}+\frac{u_\|^2}{v_{Ts}^2}(\frac{R}{L_{Ts}}-2\frac{R}{L_u})-\frac{\varpi}{n\bar\omega_{ds}}}{(2-\lambda b)f_\theta\xi+\epsilon_{\|}m\frac{\omega_b}{n\bar\omega_{ds}}-\frac{\varpi}{n\bar\omega_{ds}}+\imath \frac{\nu_{fe}(\xi,\lambda)}{n\bar\omega_{ds}}}
\end{split}\end{align}
where $\ds{\nu_{fe}=\nu_{ei}\left(\frac{v_{Te}}{\sqrt{\xi}}\right)^3Z_{\textsl{eff}}\left(\frac{\epsilon}{|1-\epsilon-\lambda|^2}\frac{0.111\delta+1.31}{11.79\delta+1}\right)}$ with $\delta=\left(\frac{|\omega|}{37.2/\epsilon Z_{\textsl{eff}}\nu_{ei}}\right)^{1/3}$ \citep{roma07}.

Now, before performing the integral over $(\xi,\lambda)$, it is worth noticing that $\mathcal{B}_1$ is odd in $k_r$. When integrating over $k_r$, it will only give a non-zero value for $\left\langle\mathcal{I}_{s,1,tr}\right\rangle$ in presence of an asymmetric eigenfunction in $k_r$. This happens only in the presence of a parallel velocity symmetry breaker\citep{peet05}: $\upar$, $\gradu$ or $\exb$ shear in \QLK{} framework. Given the fact that the Krook operator does not conserve momentum, it appears inadequate to keep this higher order term in the equation. Since $\mathcal{B}_2$ represents 5\% of $\mathcal{B}_0$ when integrated over $k_r$, higher order are not treated neither. This is why the only term actually used in \QLK{} is $m=0$. (\ref{eq:trappedm0}) therefore expresses the trapped ions functional integrated over ($\xi,\lambda$).
\begin{align}\label{eq:trappedm0}\begin{split}
\left\langle\mathcal{I}_{0,i,tr}\right\rangle_t=2f_t\int_0^1\frac{K(\kappa)\kappa}{f(\kappa)}\mathrm{d}\kappa
\Bigg[&\left(1-\frac{\upar^2}{v_{Ti}^2}\right)\left(\frac{R}{L_{Ti}}\frac{Z_2(z)}{z}+\left(\frac{R}{L_{ni}}-\frac{3}{2}\frac{R}{L_{Ti}}- z^2\right)\frac{Z_1(z)}{z}\right)\\
&-\frac{\upar}{v_{Ti}}\left(2\frac{R}{L_u}-\frac{\upar}{v_{Ti}}\frac{R}{L_{Ti}}\right)\Bigg]
\end{split}\end{align}
where $z$ is the square root of ${\frac{\varpi}{n\bar\omega_{ds}}}$ which has a positive imaginary part and $f_\kappa=2\frac{E(\kappa)}{K(\kappa)}-1+4s\left(\kappa^2-1+\frac{E(\kappa)}{K(\kappa)}\right)=\oint\frac{\d\theta}{2\pi}\frac{f_\theta}{4\sqrt{1-\lambda b}}$ with $\lambda=1-2\epsilon\kappa^2$. %For the case $m=1$, the result is similar to (\ref{eq:passing}).
Comparing (\ref{eq:trappedm0}) to (\ref{eq:passing}), the reader might have noticed that the second and third terms (lines) of (\ref{eq:passing}) are absent in (\ref{eq:trappedm0}). Indeed, the integration over $\lambda$ gives $1-2\epsilon$ for passing particles and $2\epsilon$ for trapped ions for the second term and $\frac{1}{3}$ for passing and $\frac{2}{3}f_t\epsilon$ for trapped ions for the third term. So, at lowest order in $\epsilon$, the expression for the trapped ions functional $\left\langle\mathcal{I}_{0,tr}\right\rangle$ comes down to (\ref{eq:trappedm0}). For trapped electrons, the expression (\ref{eq:I_e,tr}) is numerically integrated over $(\xi,\kappa)$.% and so are the $m=\pm1$ trapped particles terms corresponding to the second and third terms of (\ref{eq:passing}). 

The integration over $k_r$ is simplified by bounce averaging. Integration over $\theta$ being already performed for $\mathcal{I}_{0,tr}$ by bounce averaging, the only $k_r$ dependence in $\mathcal{L}_{s,tr}$ lies in $\mathcal{B}_0(k_r\delta_s)|\tilde\phi_{n\omega}|^2$ which is integrated in $k_r$ numerically. The Bessel function $\mathcal{B}_0(k_r)$ is not included in the integration above because $\rho_s\ll\delta_s)$. The expression for the trapped particle functionals: $\mathcal{L}_{0,s,tr}$ can then be written
\begin{equation}
\mathcal{L}_{0,s,tr}=\int_0^1K(\kappa)\kappa\mathcal{I}_{0,tr}\mathrm{d}\kappa\mathcal{B}_0(k_\theta\rho_s)\int\frac{\d k_r}{2\pi}\mathcal{B}_0(k_r\delta_s)|\tilde\phi_{n\omega}(k_r)|^2
\end{equation}

\section{Quasi-linear momentum flux derivation}
\label{app:mom_flux}
Using the formalism developed in Sec.~\ref{sec:gyrok} and the notations from the former appendices, the complete expression of $\Pi_\parallel$ is:
\begin{align}\label{eq:mom_flux}
\begin{split}
\Pi_{\|}=&-\sum_{\epsilon_\parallel=\pm1,s,n}\frac{n_sm_s}{B}\left(\frac{nq}{r}\right)^2\Bigg\langle\epsilon_\|\xi v_{Ts}\sqrt{1-\lambda b}e^{-\xi}\left(1+2\frac{\upar}{v_{Ts}}\epsilon_\|\sqrt{\xi(1-\lambda b)}+\frac{\upar}{v_{Ts}}^2(2\xi(1-\lambda b)-1)\right)\\
&\bigg[\frac{R\nabla n_s}{n_s}+\left(\xi-\frac{\upar}{v_{Ts}}\left(2\epsilon_\|\sqrt{\xi(1-\lambda b)}-\frac{\upar}{v_{Ts}}\right)-\frac{3}{2}\right)\frac{R\nabla T_s}{T_s}+\\
&2\left(\epsilon_\|\sqrt{\xi(1-\lambda b)}-\frac{\upar}{v_{Ts}}\right)\frac{R\nabla u_\|}{v_{Ts}}+\frac{\varpi}{n\omega_{ds}}\bigg] \Im\left(\frac{1}{\omega-n\Omega_J(\xi,\lambda)+\imath0^+}\right)\left|\tilde\phi_{n\omega}\right|^2\Bigg\rangle_{\xi,\lambda,k_r}
\end{split}
\end{align}
Apart from the saturated potential $\tilde\phi_{n\omega}$, the rest of the expression is similar to the linear gyrokinetic response presented in Sec.~\ref{sec:gyrok} except that only the imaginary part is of interest for the flux and that the integrations over ($\xi$, $\lambda$) are slightly different due to the multiplication by $v_\parallel=\pm v_{Ts}\sqrt{\xi(1-\lambda b)}$. The same techniques as before are then employed. The contributions from trapped and passing particles to the momentum flux are treated separately.
\begin{align}\label{eq:mom_flux2}\begin{split}
\Pi_{\parallel}=-\sum_{\epsilon_\parallel=\pm1,s,n}\frac{n_sm_sv_{Ts}}{B}\left(\frac{nq}{r}\right)^2\Bigg\{&\int_{-\infty}^\infty\frac{\mathrm{d}k^*}{\sqrt{\pi}}e^{-k^{*2}}\int_{-\infty}^\infty\frac{\mathrm{d}\rho^*}{\sqrt{\pi}}e^{-\rho^{*2}}\Im(\mathcal{J}_{s,pass}(k^*,\rho^*))\mathcal{B}_0(k_\theta\rho_s)\left|\tilde\phi_{n}\right|^2\\
&+\Im(\mathcal{J}_{s,tr})\int\frac{\d{k_r}}{2\pi}\mathcal{B}_0(k_\theta\rho_s)\mathcal{B}_0(k_r\delta_s)\left|\tilde\phi_n(k_r)\right|^2\Bigg\}
\end{split}\end{align}
The expression for $\mathcal{J}_{s,pass}$ is detailed in (\ref{eq:mom_flux_pass}). Its expression is very close to that of \ref{eq:I_pass}. A notable difference is that even functions ($Z1$, $Z2$, $Z3$) are replaced by odd functions ($vZ1(v)$, $vZ2(v)$, $vZ3(v)$). This indicates that without rotation the momentum is zero.
{\small
\begin{align}\label{eq:mom_flux_pass}\begin{split}
\mathcal{J}_{s,pass}=&\frac{2}{f_\theta}\bigg[\frac{R}{L_{Ts}}\frac{V_+Z_2(V_+)-V_-Z_2(V_-)}{V_+-V_-}+\left(\frac{R}{L_{ns}}-\frac{3}{2}\frac{R}{L_{Ts}}-\frac{\varpi}{n\bar\omega_{ds}}\right)\frac{V_+Z_1(V_+)-V_-Z_1(V_-)}{V_+-V_-}\bigg]\\
+\frac{4}{3f_\theta}&\bigg[\uparns \frac{R}{L_{Ts}}\frac{Z_3(V_+)-Z_3(V_-)}{V_+-V_-}+
\left(\frac{R}{L_u}+\uparns\left(\frac{R}{L_{ns}}-\frac{5}{2}\frac{R}{L_{Ts}}-\frac{\varpi}{n\bar\omega_{ds}}\right)\right)\frac{Z_2(V_+)-Z_2(V_-)}{V_+-V_-}\bigg]\\
+\frac{\upar}{f_\theta v_{Ts}}&\bigg[\uparns \frac{R}{L_{Ts}}\frac{V_+Z_3(V_+)-V_-Z_3(V_-)}{V_+-V_-}+
\left(2\frac{R}{L_u}+\uparns\left(\frac{R}{L_{ns}}-\frac{7}{2}\frac{R}{L_{Ts}}-\frac{\varpi}{n\bar\omega_{ds}}\right)\right)\frac{V_+Z_2(V_+)-V_-Z_2(V_-)}{V_+-V_-}\bigg]\\
-\frac{2\upar}{f_\theta v_{Ts})}&\left[\uparns \frac{R}{L_{Ts}}\frac{V_+Z_2(V_+)-V_-Z_2(V_-)}{V_+-V_-}+\left(2\frac{R}{L_u}+\uparns\left(\frac{R}{L_{ns}}-\frac{5}{2}\frac{R}{L_{Ts}}-\frac{\varpi}{n\bar\omega_{ds}}\right)\right)\frac{V_+Z_1(V_+)-V_-Z_1(V_-)}{V_+-V_-}\right]
\end{split}\end{align}
}
For trapped particles, there is no contribution to the momentum flux at lowest order in $\epsilon$ because the functional is odd in $\xi$ due to the multiplication by $v_\parallel$ of the linear response. However, when expanding up to first order in $\sqrt{\epsilon}$, there is a net contribution from trapped particles, detailed in (\ref{eq:mom_flux_tr}).
\begin{equation}\label{eq:mom_flux_tr}
\mathcal{J}_{s,tr}=2\bar\omega_b\Bigg[\left(\frac{R}{L_u}+\uparns\left(\frac{R}{L_{ns}}-\frac{5}{2}\frac{R}{L_{Ts}}-\frac{\varpi}{n\bar\omega_{ds}}\right)\right)\frac{Z_2(z)}{z}+\frac{\upar}{v_{Ts}}\frac{R}{L_{Ts}}\frac{Z_3(z)}{z}\Bigg]
\end{equation}

\section{Fluid model derivation}
\label{app:fluid}
The fluid limit approximation consists in considering events sufficiently fast decorrelated by collisions such that $\varpi=\omega-n\omega_{E\times B}\gg\bar\omega_{di}$ and $\varpi\gg\kpar v_{\parallel i}$. This approximation enables the resonance to be developed in power of the small quantities $\frac{\omega_{ds}}{\varpi}$, $\frac{\kpar \vpar}{\varpi}$ and obtain a polynomial expression in $\varpi$ as detailed in (\ref{eq:electroneutralite4}).

For short wavelengths: $k_\bot\rho_i<1$, the Pade approximation is performed: $\mathcal{B}_0(k_\bot\rho_i)\approx 1-\frac{k_\bot^2\rho_i^2}{2}$. At this spatial scale events are sufficiently slow such that $\omega\ll \kpar v_{\parallel e}$. Passing electrons are then considered adiabatic. In contrast, TEM space and time scales being the same as ions modes, trapped electrons are treated by the model. Since $k_r\delta_e<k_r\rho_i<1$, the Bessel functions on trapped electrons are considered close to unity $\mathcal{B}_0(k_r\delta_e)\approx 1$. For trapped ions, the finite banana width effects are expended in power of $k_r$ too: $ \mathcal{B}_0(k_r\delta_i)\approx 1-\frac{k_r^2\delta_i^2}{2}$. The resulting expression for the eigenmode is given in (\ref{eq:electroneutralite4}). 
\begin {align}
\begin {split}\label{eq:electroneutralite4}
\Bigg[\frac{n_e}{T_e}&\left(\left\langle1-\left(1-\frac{n\omega_e^*}{\varpi}\right)\left(1+\frac{n\omega_{de}}{\varpi}\right)\right\rangle_t+f_p\right)+\\
\sum_i\frac{n_iZ_i^2}{T_i}&\left\langle\left(1-\left(1-\frac{n\omega_i^*}{\varpi}\right)\left(1+\frac{n\omega_{di}}{\varpi}\right)\right) \left(1-\frac{k_r^2\delta_i^2}{4}\right)\right\rangle_t+\\
\sum_i\frac{n_iZ_i^2}{T_i}&\left\langle\left(1-\left(1-\frac{n\omega_i^*}{\varpi}\right) \left(1+\frac{n\omega_{di}}{\varpi}+\frac{k_{\|}v_{\|i}}{\varpi}+\frac{k_\|^2v_{\|i}^2}{\varpi^2}\right)\right)\left(1-\frac{k_\bot^2\rho_i^2}{2}\right)\right\rangle_{p}\Bigg]\tilde\phi=0
\end {split}
\end{align}
The integration over $k_r$ present in (\ref{eq:electroneutralite2}) is not performed in (\ref{eq:electroneutralite4}) since $\theta=k_rd$ and $\theta$ is a parallel coordinate label in the ballooning representation. The goal of the model being to capture the radial and parallel variations of the eigenfunction, capturing the dependence on $k_r$ is crucial. This is done through an inverse Fourier transform from $k_r$ to $x$. But first, (\ref{eq:electroneutralite4}) is simplified by using the electroneutrality condition $\sum_in_iZ_i^2=n_e$. To simplify (\ref{eq:electroneutralite4}), new quantities are defined: $\ceff=\sqrt{\frac{T_e}{m_p}}$ is an effective thermal velocity, $\delta_{eff}^2=\frac{3}{4}(1+\frac{f_t}{f_p}\frac{q^2}{4\epsilon})\frac{4m_pT_e}{e^2B^2}$ represents both finite ion Larmor radius and banana width effects. Finally $\tau=T_i/T_e$. Moreover, considering the low Mach number limit, only the terms linear in $\uparns$ are kept.
\begin{align}
\label{eq:limite_fluide}
\begin{split}
\frac{n_e}{T_e}\Bigg[&f_p\left(1-\frac{n\omega_{ne}^*}{\varpi}+\left(\frac{2n\omega_{d}}{\varpi}+\frac{k_\theta^2\rho_{\textsl{eff}}^2}{2}+\frac{k_r^2\delta_{\textsl{eff}}^2}{2}-\frac{\kpar^2\ceff^2}{2\varpi^2}\right)\left(1+\frac{n\omega_{pi}^*}{\varpi}\right)\right)-\\
&f_p\left(\frac{n\omeu}{\varpi}+\frac{\upar}{\ceff}\left(\frac{Z_{\textsl{eff}}}{\tau}+\frac{n\omega_{ne}^*}{\varpi}-\frac{8n\bar\omega_{d}}{\varpi}\right)\right)\frac{\kpar\ceff}{\varpi}+\frac{f_t}{2}\frac{n\bar\omega_{d}n\omega_{pe}^*}{\varpi^2}\Bigg]\tilde\phi=0
\end{split}
\end{align}
The passing particle curvature drift reads: $n\omega_d=n\bar\omega_d(\cos(k_rd)+(\hat{s}k_rd-\alpha\sin(k_rd))\sin(k_rd))$ since $\theta=k_rd$. $n\bar\omega_d=n\bar\omega_{de}=-1/\tau n\bar\omega_{di}$. As the ITG turbulence exhibits ballooned modes around $\theta=0$ \citep{brunner98} (which was used for our ballooning representation simplification \citep{bour02}), the following linearization is possible: $n\omega_d\rightarrow n\bar\omega_d(1+1+(k_rd)^2(\hat{s}-\alpha-0.5))$ \citep[see][App. A]{roma07}. After this operation, (\ref{eq:limite_fluide}) is finally polynomial in $k_r$. The inverse Fourier transform in $k_r$ is then performed. The structure of a second order differential equation becomes clear as $k_r$ is transformed into $-\imath\dx$. 

(\ref{eq:limite_fluide}) is multiplied by $\varpi^2$ and $\varpi$ is replaced by $\omega-n\omega_{E\times B}$ to make the $x$ dependence of $\varpi$ appear. Indeed, a radial dependence in $\omega_{E\times B}$ is taken into account. The radial electric field is considered smooth enough such that it can be linearized into $E_r\rightarrow E_{r0}+E_r'x+ O(x^2)$ with $x=r-r_0$ being a small parameter. Therefore, $n\omega_{E\times B}=\frac{\kthe E_r}{B}\rightarrow \frac{\kthe E_{r0}}{B}+\frac{\kthe E_{r}'}{B}x+ O(x^2)=n\omega_{E0}+\kthe\gamma_E x+O(x^2)$. $\omega$ considered below is $\omega-n\omega_{E0}$ since this Doppler shift does not modify the stability of the mode. 
\begin{align}\label{eq:lim_fluide1}\begin{split}
\Bigg[&\left(\omega\left(\frac{\deff^2}{2}\ddx-\frac{\kthe^2\reff^2}{2}\right) -2n\bar\omega_{d}+\frac{\kpar^{\prime 2}\ceff^2}{2\omega}x^2\right)\left(\omega-\kthe\gamma_E x-n\omepi\right)
-\frac{f_t}{f_p}n\omepe n\bar\omega_{d}\\
&-(\omega-\kthe\gamma_E x)\left(\omega-\kthe\gamma_E x-n\omen\right)+\kpar^\prime\ceff\left(n\omeu+\frac{\upar}{\ceff}\left(\frac{\zeff}{\tau}\omega+n\omen-8n\bar\omega_{d}\right)\right)x\Bigg]\tilde\phi=0
\end{split}\end{align}
$\deff$ is defined as $\deff=\delta_{eff}^2+4\frac{n\bar\omega_d}{\omega}(\hat{s}-\alpha-0.5)d^2$, containing all terms proportional to $k_r^2$. (\ref{eq:lim_fluide1}) is not linear and there is no general analytic solution of it. But, the ballooning representation used to derive the gyrokinetic dispersion relation (\ref{eq:lin_gyroK}) assumes a ballooned turbulence around $\theta=0$. This is not correct if $\gamma_E\gg\omega$. $x$ being small, any term in $k_\bot x$ and $x^3$ or superior are neglected. This results in the following second order linear differential equation:
\begin{align}\label{eq:lim_fluide3}\begin{split}
\Bigg[&\left(\omega\left(\frac{\deff^2}{2}\ddx-\frac{\kthe^2\reff^2}{2}\right) +\frac{\kpar^{\prime 2}\ceff^2}{2\omega}x^2\right)\left(\omega-n\omepi\right) -2n\bar\omega_{d}(\omega-\kthe\gamma_E)
-\omega^2+2\kthe\gamma_E+\\&\left(\omega-\kthe\gamma_E\right)n\omen-\frac{f_t}{f_p}n\omepe n\bar\omega_{d}+ \kpar^\prime\ceff\left(n\omeu+\frac{\upar}{\ceff}\left(\frac{\zeff}{\tau}\omega+n\omen-8n\bar\omega_{d}\right)\right)x\Bigg]\tilde\phi=0
\end{split}\end{align}

\section{Eigenmodes in strong TEM cases}
\label{app:TEM}
\begin{figure*}
		\subfigure{\includegraphics[width=0.5\textwidth]{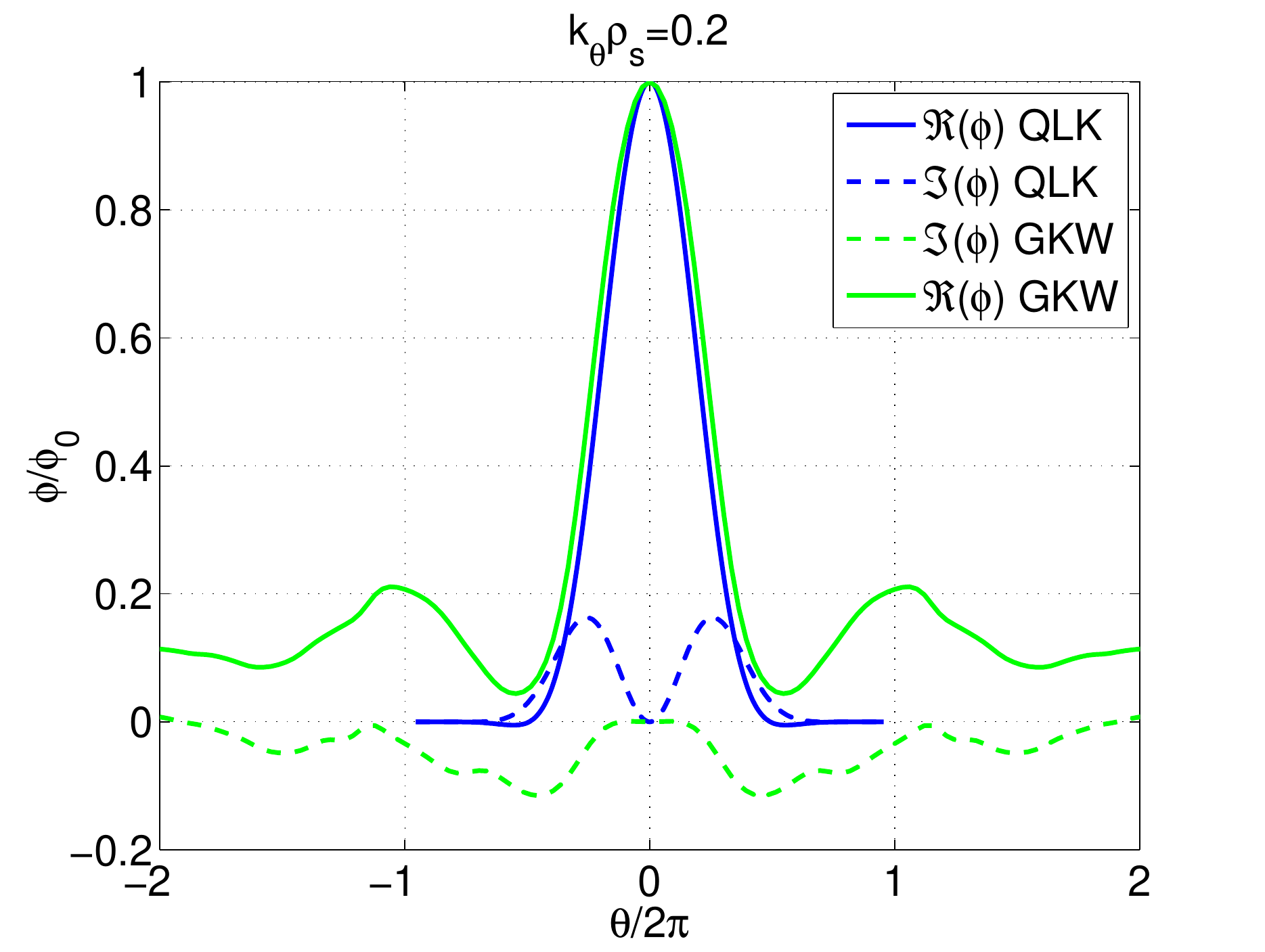}}
		\subfigure{\includegraphics[width=0.5\textwidth]{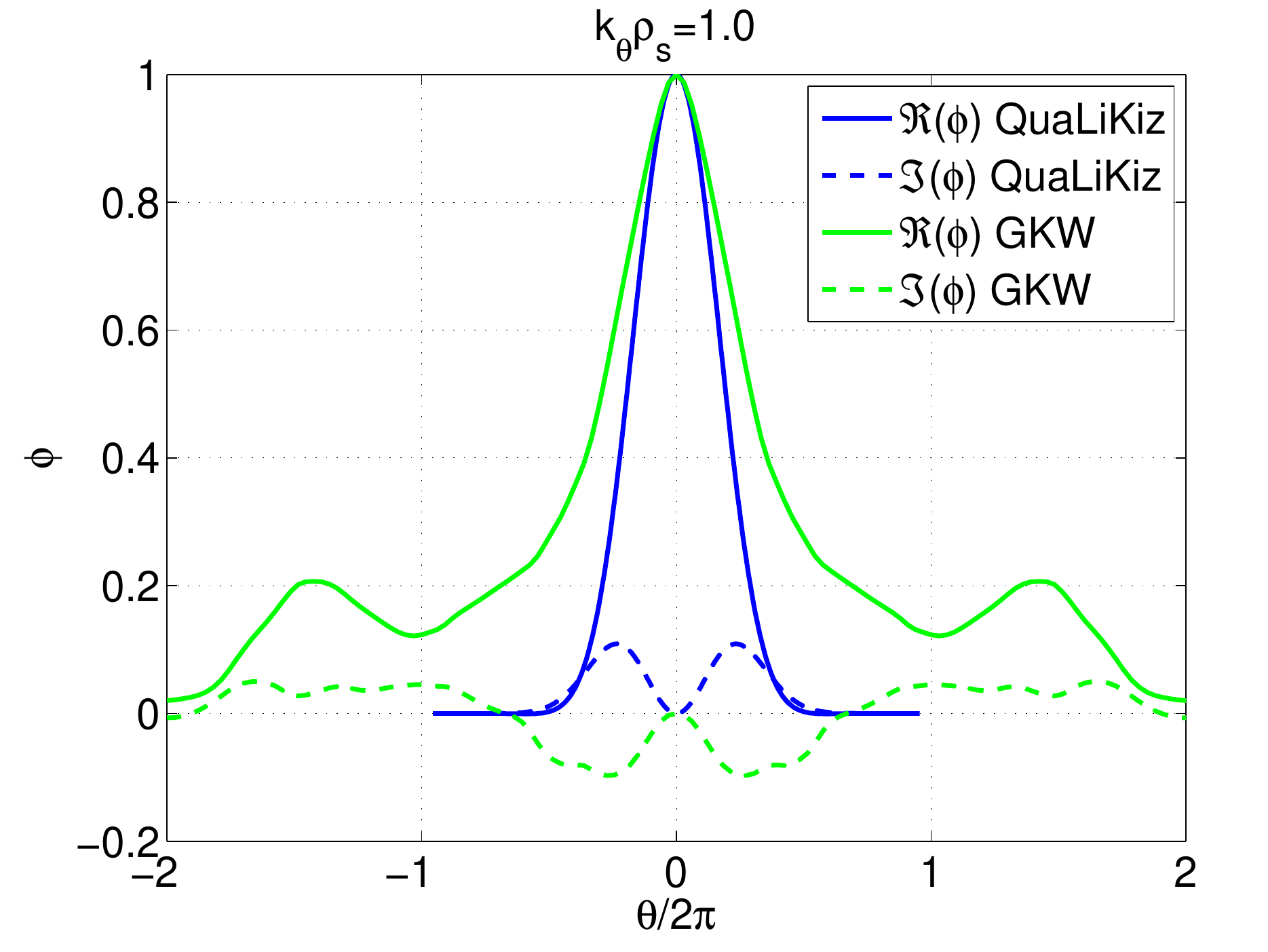}}
		\caption{\label{fig:comp_phi_TEM}Parallel structure of the eigenfunctions showing the increased $\theta$ spreading with $\kthe\rho_s$ in the case of TEM. $R/L_{Ti}=0$, other parameters from GA-std test case. $\kthe\rho_s=0.2$ left panel. $\kthe\rho_s=1.0$ right panel.}
\end{figure*}

A GA-std case with $R/L_{Ti}=0$ keeping $R/L_{Te}=9$ is studied; in this case TEM are strongly dominant. In Figure~\ref{fig:comp_phi_TEM}, \QLK{} eigenmodes are compared to \GKW{} for two poloidal wave numbers values: $\kthe\rho_s=0.2$ and $\kthe\rho_s=1.0$.
As foreseen, looking at the real part of the eigenmodes, \GKW{} ones extent over a large domain $|\theta|>\pi$ which is not captured by our fluid model. In contrast, the agreement is satisfactory for $\theta$ inside $[-\pi; \pi]$, which is consistent with the restriction made in \QLK{} in the ballooning representation. However, \GKW{} $\tilde\phi$ imaginary part flips sign between ITG and TEM whereas \QLK{} one remains positive. Finally, inside $[-\pi; \pi]$, the agreement between \QLK{} and \GKW{} is better at lower $k_\theta\rho_s$ as expected due to the linearization of the Bessel functions in the fluid model. This is important since $k_\theta\rho_s\approx0.2$ corresponds to the spatial scales responsible for most of the transport. Overall, in cases where TEM are strongly dominant, it can be foreseen that the growth rates predicted by \QLK{} will be underestimated compared to self-consistent gyrokinetic simulations and this underestimation will increase with increasing $\kthe\rho_s$.
%Indeed, for flat density conditions, $R/L_n=0$, and overall other same parameters, \QLK{} $\Im(\tilde\phi)$ flips sign for higher $k_\theta\rho_s$ value but not at low $k_\theta\rho_s$ as plotted on Figure \ref{fig:comp_phi_TEM_An0}.This means that $\Im(\tilde\phi)$ does not have the correct dependence with $R/L_n$ nor $k_\theta\rho_s$ for TEM cases.
Nevertheless, the ion temperature gradients are never zero in experimental cases. Thus realistic eigenfunctions are generally well reproduced by the fluid model used in \QLK. 

\bibliography{Biblio_new_new}

%\addcontentsline{toc}{chapter}{Bibliographie}
\end{document}